\documentclass[bibyear]{aa}
\usepackage{natbib}
\usepackage{longtable}
\usepackage{graphicx}
\usepackage{txfonts}
\def\tyc{\mbox{TYC\,2187-512-1}}
\input cyracc.def
\font\tencyr=wncyr8
\def\cyr{\tencyr\cyracc}

\begin{document}

   \title{The CARMENES search for exoplanets around M dwarfs}

   \subtitle{Two Saturn-mass planets orbiting active stars\thanks{The CARMENES radial-velocity data are only available in electronic form at the CDS via anonymous ftp to cdsarc.u-strasbg.fr (130.79.128.5) or via http://cdsweb.u-strasbg.fr/cgi-bin/qcat?J/A+A/}}

   \author{A.~Quirrenbach
          \inst{1}
          \and
          V.\,M.~Passegger
          \inst{2,3}
          \and
          T.~Trifonov
          \inst{4,5}
          \and
          P.\,J.~Amado
          \inst{6}
          \and
          J.\,A.~Caballero
          \inst{7}
          \and
          A.~Reiners
          \inst{8}
          \and
          I.~Ribas
          \inst{9,10}
          \and
          J.~Aceituno
          \inst{6,11}
          \and
          V.\,J.\,S.~B\'ejar
          \inst{12,13}
          \and
          P.~Chaturvedi
          \inst{14}
          \and
          L.~Gonz\'alez-Cuesta
          \inst{12,13}
          \and
          T.~Henning
          \inst{4}
          \and
          E.~Herrero
          \inst{9,10}
          \and
          A.~Kaminski
          \inst{1}
          \and
          M.~K\"urster
          \inst{4}
          \and
          S.~Lalitha
          \inst{15}
          \and
          N.\,Lodieu
          \inst{12,13}
          \and
          M.\,J.~L\'opez-Gonz\'alez
          \inst{6}
          \and
          D.~Montes
          \inst{16}
          \and
          E.~Pall\'e{}
          \inst{12,13}
          \and
          M.~Perger
          \inst{9,10}
          \and
          D.~Pollacco
          \inst{17}
          \and
          S.~Reffert
          \inst{1}
          \and
          E.~Rodr\'iguez
          \inst{6}
          \and
          C.~Rodr\'iguez L\'opez
          \inst{6}
          \and
          Y.~Shan
          \inst{8}
          \and
          L.~Tal-Or
          \inst{18,8}
          \and
          M.\,R.~Zapatero~Osorio
          \inst{19}
          \and
          M.~Zechmeister
          \inst{8}
          }

   \institute{Landessternwarte, Zentrum f\"ur Astronomie der Universit\"at Heidelberg, K\"onigstuhl 12, D-69117 Heidelberg, Germany
         \and
              Hamburger Sternwarte, Gojenbergsweg 112, D-21029 Hamburg, Germany
         \and
              Homer L. Dodge Department of Physics and Astronomy, University of Oklahoma, 440 West Brooks Street, Norman, OK 73019, USA
         \and
              Max-Planck-Institut f\"ur Astronomie, K\"onigstuhl 17, D-69117 Heidelberg, Germany
         \and
              Department of Astronomy, Sofia University ``St Kliment Ohridski'', 5 James Bourchier Blvd, BG-1164 Sofia, Bulgaria
         \and
              Instituto de Astrof\'{\i}sica de Andaluc\'{\i}a (CSIC), Glorieta de la Astronom\'{\i}a s/n, E-18008 Granada, Spain
         \and
              Centro de Astrobiolog\'{\i}a (CSIC-INTA), ESAC, Camino Bajo del Castillo s/n, E-28692 Villanueva de la Ca\~nada, Madrid, Spain
         \and
              Institut f\"ur Astrophysik, Georg-August-Universit\"at G\"ottingen, Friedrich-Hund-Platz 1, D-37077 G\"ottingen, Germany
         \and
              Institut de Ci\`encies de l'Espai (ICE, CSIC), Campus UAB, c/ de Can Magrans s/n, E-08193 Bellaterra, Barcelona, Spain
         \and
              Institut d'Estudis Espacials de Catalunya (IEEC), E-08034 Barcelona, Spain
         \and
              Centro Astron\'omico Hispano-Alem\'an (MPG-CSIC), Observatorio Astron\'omico de Calar Alto, Sierra de los Filabres, E-04550 G\'ergal, Almer\'{\i}a, Spain
         \and
              Instituto de Astrof\'{\i}sica de Canarias, c/ V\'{\i}a L\'actea s/n, E-38205 La Laguna, Tenerife, Spain
         \and
              Departamento de Astrof\'{\i}sica, Universidad de La Laguna, E-38206 Tenerife, Spain
         \and
              Th\"uringer Landessternwarte Tautenburg, Sternwarte 5, D-07778 Tautenburg, Germany
         \and
              School of Physics \& Astronomy, University of Birmingham, Edgbaston, Birmingham B15 2TT, United Kingdom
         \and
              Departamento de F\'isica de la Tierra y Astrof\'{\i}sica \& IPARCOS-UCM (Instituto de F\'isica de Part\'iculas y del Cosmos de la UCM), Facultad de Ciencias F\'{\i}sicas, Universidad Complutense de Madrid, E-28040 Madrid, Spain
         \and
              Department of Physics, University of Warwick, Coventry CV4 7AL, United Kingdom
         \and
              Department of Physics, Ariel University, Ariel 40700, Israel
         \and
              Centro de Astrobiolog\'{\i}a (CSIC-INTA), Campus INTA, Carretera de Ajalvir km 4, E-28850 Torrej\'on de Ardoz, Madrid, Spain
             }

   \date{Received 14 December 2021 / accepted 15 March 2022}
  \abstract
  {The CARMENES radial-velocity survey is currently searching for planets in a sample of 387 M dwarfs.
Here we report on two Saturn-mass planets orbiting \tyc{} ($M_\star = 0.50\,M_\odot$) and TZ\,Ari ($M_\star = 0.15\,M_\odot$), respectively. We obtained supplementary photometric time series, which we use along with spectroscopic information to determine the rotation periods of the two stars. In both cases, the radial velocities also show strong modulations at the respective rotation period. We thus modeled the radial velocities as a Keplerian orbit plus a Gaussian process representing the stellar variability. \tyc{} is found to harbor a planet with a minimum mass of 0.33\,$M_{\rm Jup}$ in a near-circular 692-day orbit. The companion of TZ\,Ari has a minimum mass of 0.21\,$M_{\rm Jup}$, orbital period of 771\,d, and orbital eccentricity of 0.46. We provide an overview of all known giant planets in the CARMENES sample, from which we infer an occurrence rate of giant planets orbiting M dwarfs with periods up to 2 years in the range between 2\,\% to 6\,\%. TZ\,Ari\,b is only the second giant planet discovered orbiting a host with mass less than 0.3\,$M_\odot$. These objects occupy an extreme location in the planet mass versus host mass plane. It is difficult to explain their formation in core-accretion scenarios, so they may possibly have been formed through a disk fragmentation process.}


\titlerunning{Two Saturn-mass planets around M dwarfs}
\authorrunning{A. Quirrenbach et al.}

   \maketitle
\section{Introduction}
\label{Introduction}

The occurrence rate of giant planets depends on the properties of their host stars. This can be understood in the context of their formation and early migration in the circumstellar disk. As a general trend, the dust masses in protoplanetary disks increase with the mass of their host \citep{Manara2018}. Likewise, the number of gas giants in orbits with periods up to a few years increases with stellar mass \citep{Johnson2010a}, with an apparent maximum near 2\,$M_{\odot}$ and a sharp decline towards even higher masses \citep{Reffert2015}. Surveys for planets orbiting M dwarfs play an important role in this context, as this spectral type spans nearly an order of magnitude in mass, from the hydrogen burning limit at 0.07\,$M_{\odot}$ to about 0.6\,$M_{\odot}$. While the first planet found around an M dwarf, GJ\,876\,b, is surprisingly massive with $m = 2.1\,M_{\rm Jup}$ \citep{Marcy1998, Delfosse1998, Trifonov2018}, gas giants orbiting M dwarfs are actually quite rare \citep{Bonfils2013,Sabotta2021}. In a similar vein, a trend of fewer large planets with radii of $R \gtrsim 1.4\,R_\oplus$ is found towards decreasing stellar temperatures \citep{Dressing2013}.

Giant planets orbiting low-mass stars contain important information about planet formation processes, as they challenge the core accretion paradigm and, specifically, pebble accretion as the dominant mechanism of core growth \citep{Morales2019}. Here, we report on the discovery of two Saturn-mass planets from the CARMENES survey, which is currently carrying out precise radial-velocity (RV) observations of a sample of 387 M dwarfs \citep[for a recent overview, see][]{Quirrenbach2020}. As both host stars \tyc{} and TZ\,Ari also show clear RV variations at the stellar rotation period, we modeled these as Gaussian processes (GPs) alongside the Keplerian signals.

Section~\ref{Targets} contains the fundamental parameters and other pertinent information about the two target stars. In Sect.~\ref{Photometry}, we describe supplementary photometry carried out at several observatories. The CARMENES observations of both stars, and RVs of TZ\,Ari gathered from the HARPS and HIRES archives, are presented in Sect.~\ref{Spec}. Photometric and spectroscopic information is combined in Sect.~\ref{Rot} to determine the stellar rotation periods. In Sect.~\ref{Modeling}, we introduce our GP model, apply it to the RV time series, and determine the planetary parameters. We discuss our results in Sect.~\ref{Discussion} and conclude with a summary in Sect.~\ref{Conclusions}.

\section{Target stars}
\label{Targets}

Table~\ref{tab:star_props} provides a summary of the stellar properties of both stars. In this section, we provide a brief description of how they were derived.

\subsection{\tyc{}}

The star \tyc{} (Karmn J21221+229\footnote{CARMENES identifier from the Carmencita database, see \citet{Caballero2016a}.}) is an early M dwarf with spectral type M1.0\,V \citep{Lepine2013}. We adopted the distance of 15.485 $\pm$ 0.005\,pc from $Gaia$ EDR3 \citep[$\pi =$ 64.580 $\pm$ 0.021 mas,][]{Gaia2021}. The photospheric parameters effective temperature, $T_{\rm eff}$, surface gravity, $\log{g}$, and metallicity, [Fe/H], were determined by \citet{Passegger2019}. These authors used updated PHOENIX models with the latest atomic and molecular line lists, updated Solar abundances \citep{Caffau2011}, and a new equation of state \citep{Meyer2017}, and fitted them to high-resolution CARMENES spectra in the visible and near-infrared wavelength ranges. During this process, a rotational velocity of $v \sin{i_\star} = 2$\,km\,s$^{-1}$ was assumed, which was reported as the upper limit for this star by \citet{Reiners2018}. A recent analysis with the {\tt SteParSyn} code by \citet{Marfil2021} yielded systematically lower values of [Fe/H]; the difference of about 0.2\,dex is indicative of persistent difficulties in determining accurate parameters for M dwarfs. The bolometric luminosity, $L,$ was taken from \citet{Schweitzer2019}. The spectral energy distribution was integrated from a number of apparent magnitudes in broad band passes over the whole spectral range from the blue optical to the mid-infrared,
as described by \citet{Cifuentes2020}. Together with the $Gaia$ distance, the luminosity, $L,$ can be derived. The stellar radius was calculated from $T_{\rm eff}$ and $L$ via the Stefan-Boltzmann law. With this, we derived the stellar mass from the mass-radius relation of \citet{Schweitzer2019}. The pseudo-equivalent width pEW(H$\alpha$) is $+0.4 \pm 0.1$\,\AA{} \citep{Jeffers2018,Lepine2013,Gaidos2014}\footnote{Note that there is a sign error in the equivalent widths given by \cite{Gaidos2014} and in the associated electronic tables.}. A slightly different quantity, pEW$'$(H$\alpha$), is given by \citet{Schoefer2019}. It is defined by subtracting the spectrum of an inactive star before computing the pEW, thus isolating the emission component of the line. The value for \tyc{} is $-0.032 \pm 0.006$\,\AA{}. From our CARMENES spectra, we find the star to be moderately active, with clear signals in the activity indicators, but no H$\alpha$ emission in the individual spectra.

\subsection{TZ\,Ari}
\label{TZprop}

Different spectral types have been given in the literature for the M dwarf TZ\,Ari (Karmn J02002+130, Gl\,83.1). \citet{Lepine2013} claimed the star to be an M5.0\,V, whereas \cite{Alonso-Floriano2015} stated it was an M3.5:\,V. Considering $T_{\rm eff}$, measured by \citet{Passegger2019} (spectroscopic, $3154 \pm 54$\,K) and \cite{Houdebine2019} (photometric, $3158 \pm 23$\,K), a spectral type of M5.0\,V seems to be a more accurate assumption. TZ\,Ari is very close to the Sun, with a distance of only $4.470 \pm 0.001$\,pc \citep[$\pi = 223.73 \pm 0.07$\,mas,][]{Gaia2021}. The bolometric luminosity as well as the stellar radius and mass were calculated in the same way as for \tyc. The parameters $T_{\rm eff}$, $\log{g}$, and [Fe/H] were taken from \citet{Passegger2019}. Regarding the projected rotational velocity, \citet{Reiners2018} measured an upper limit of 2\,km\,s$^{-1}$. There is no rotational period in the literature for this star (but see Sect.~\ref{Rot}); from $v\sin{i_\star}$ and the stellar radius we can estimate a lower limit of $P_{\rm rot} / \sin{i_\star} > 4.15$\,d. This M dwarf is a flare star \citep{Kunkel1968, Rodriguez_Martinez2020}, hence, its variable star designation. It shows significant H$\alpha$ emission with pEW(H$\alpha$) measurements yielding $-2.056 \pm 0.012$\,\AA{} \citep{Jeffers2018} or $-1.981 \pm 0.030$\,\AA{} \citep{Newton2017} and pEW$'$(H$\alpha$)~$ = -1.906 \pm 0.025$\,\AA{} \citep{Schoefer2019}.

\subsection{Kinematics and ages}
\label{Kin}

Kinematically, both stars are members of the thin disk, without any obvious membership in a moving group or association \citep{Cortes-Contreras2020}. With a rotation period of 40\,d (see Sect.~\ref{Rot}), \tyc{} is most likely older than 2.5\,Gyr. TZ\,Ari is a much faster rotator, but a comparison with M dwarfs of known ages \citep{Popinchalk2021} indicates that its age is no younger than that of the Hyades, 750\,Myr.

We also queried the Gaia EDR3 database for wide co-moving companions within a search radius of $2^\circ$, but did not find any objects with a matching parallax.

\begin{table*}
\centering
\small
\caption{Stellar parameters of \tyc{} and TZ\,Ari.} \label{tab:star_props}
\begin{tabular}{lccr}
\hline\hline
\noalign{\smallskip}
Parameter   & \multicolumn{2}{c}{Value}             & Reference \\
\hline
\noalign{\smallskip}
\multicolumn{4}{c}{\em Name and identifiers}\\
\noalign{\smallskip}
Name    & \tyc{} & TZ\,Ari & \\
Gliese  & -- & 83.1    & G69 \\
Karmn   & J21221+229 & J02002+130  & AF15 \\
\noalign{\smallskip}
\multicolumn{4}{c}{\em Coordinates and spectral type}\\
\noalign{\smallskip}
$\alpha$ & 21:22:06.41 & 02:00:14.16 & {\it Gaia} EDR3\tablefootmark{(a)} \\
$\delta$ & +22:55:55.0 & +13:02:38.7  & {\it Gaia} EDR3\tablefootmark{(a)} \\
Sp.type & M1.0\,V & M5.0\,V & Le13\\
        & -- & M3.5:\,V & AF15\\
$G$ [mag] & $9.7752 \pm 0.0028$ & $10.6811 \pm 0.0029$ & {\it Gaia} EDR3 \\
$J$ [mag] & $7.400 \pm 0.018$ & $7.514 \pm 0.017$ & 2MASS\\
\noalign{\smallskip}
\multicolumn{4}{c}{\em Parallax and kinematics}\\
\noalign{\smallskip}
$\varpi$ [mas] & $64.580 \pm 0.021$ & $223.732 \pm 0.070$ & {\it Gaia} EDR3 \\
$d$ [pc] & $15.485 \pm 0.005$ & $4.470 \pm 0.001$ & {\it Gaia} EDR3\\
$\mu_{\alpha}\cos\delta$ [$\mathrm{mas\,yr^{-1}}$] & $+108.77 \pm 0.02$ & $+1096.46 \pm 0.07$ & {\it Gaia} EDR3 \\
$\mu_{\delta}$ [$\mathrm{mas\,yr^{-1}}$] & $+117.79 \pm 0.02$ & $-1771.53 \pm 0.06$ & {\it Gaia} EDR3 \\
$\gamma$ [$\mathrm{km\,s^{-1}}$] & $5.353 \pm 0.023$ & $-28.832 \pm 0.025$ & La20 \\
$U$ [$\mathrm{km\,s^{-1}}$] & $-9.91$ & $+13.77$ & This work\tablefootmark{(b)} \\
$V$ [$\mathrm{km\,s^{-1}}$] & $+7.73$ & $-51.14$ & This work\tablefootmark{(b)} \\
$W$ [$\mathrm{km\,s^{-1}}$] & $-1.69$ & $+3.54$ & This work\tablefootmark{(b)} \\
Gal. population & Thin disk & Thin disk & This work \\
\noalign{\smallskip}
\multicolumn{4}{c}{\em Photospheric parameters}\\
\noalign{\smallskip}
$T_{\rm eff}$ [K] & $3734 \pm 54$ & $3154 \pm 54$ & Pa19\\
$\log{g}$ [dex] & $4.68 \pm 0.06$ & $5.01 \pm 0.06$ & Pa19\\
$[$Fe/H$]$ [dex] & $+0.08 \pm 0.19$ & $-0.13 \pm 0.19$ & Pa19\\
                 & $-0.18 \pm 0.08$ & $-0.25 \pm 0.14$ & Ma21\\
$v \sin i_\star$ [$\mathrm{km\,s^{-1}}$] & < 2 & < 2 & R18\\
\noalign{\smallskip}
\multicolumn{4}{c}{\em Physical parameters}\\
\noalign{\smallskip}
$M_\star$ [$M_{\odot}$] & $0.498 \pm 0.019$ & $0.150 \pm 0.010$ & Schw19\\
$R_\star$ [$R_{\odot}$] & $0.495 \pm 0.014$ & $0.164 \pm 0.005$ & Schw19\\
$L_\star$ [$L_\odot$] &  $0.0416 \pm 0.0007$ & $0.00254 \pm 0.00002$ & Schw19\\
\noalign{\smallskip}
\multicolumn{4}{c}{\em Activity and rotation}\\
\noalign{\smallskip}
pEW$'$(H$\alpha$) [\AA{}] & $-0.032 \pm 0.006$ & $-1.906 \pm 0.025$ & Schf19\tablefootmark{(c)}\\
$S_{\rm MWO}$        & -- & 15.8 & BS18 \\
                     & -- & $6.2 \pm 1.5$ & AD17\\
$\log R'_{\rm HK}$   & -- & $-4.16$ & BS18 \\
                     & -- & $-4.79 \pm 0.42$ & AD17\\
                     & $-4.847 \pm 0.015$ & $-4.46$\tablefootmark{(d)} & Pe21\\
$P_{\rm rot}$ [d]    & 41 $\pm$ 1.7 & -- & DA19\\
                     & 40 $\pm$ 1   & 1.96 $\pm$ 0.02 & This work\\
\noalign{\smallskip}
\hline
\end{tabular}
\tablefoot{
\tablefoottext{a}{Gaia EDR3 coordinates are given in the ICRS at epoch J2016.0.}
\tablefoottext{b}{For the computation of $U$, $V$, and $W$, a correction for the gravitational redshift was applied to the observed radial velocity $\gamma$, see La20.}
\tablefoottext{c}{pEW$'$(H$\alpha$) is the pseudo-equivalent width of the line after subtraction of the spectrum of an inactive reference star.}
\tablefoottext{d}{Median value of time series tabulated by Pe21.}
}
\tablebib{
    G69: \cite{Gliese1969};
    AF15: \cite{Alonso-Floriano2015};
    {\it Gaia} EDR3: \citet{Gaia2021};
    Le13: \citet{Lepine2013};
    2MASS: \cite{Skrutskie2006};
    La20: \cite{Lafarga2020};
    Pa19: \cite{Passegger2019};
    Ma21: \cite{Marfil2021};
    R18: \cite{Reiners2018};
    Schw19: \cite{Schweitzer2019};
    Schf19: \cite{Schoefer2019};
    BS18: \cite{Boro_Saikia2018};
    AD17: \cite{Astudillo-Defru2017};
    Pe21: \cite{Perdelwitz2021};
    DA19: \cite{Diez_Alonso2019}.
}
\end{table*}

\section{Photometry}
\label{Photometry}

\subsection{SuperWASP}

The Super-Wide Angle Search for Planets (SuperWASP) Survey \citep{Pollacco2006} comprises two robotic eight-camera telescope arrays stationed in La Palma, Spain and in Sutherland, South Africa. The cameras each have a CCD with
a field size of $\sim 60$\,deg$^2$, giving a plate scale of $13\farcs7$ pixel$^{-1}$. They are fitted with broadband filters spanning 400 to 700\,nm. In operation since 2004, these systems monitor the entire sky at high cadence and deliver good quality photometry (i.e., characteristic precision $\lesssim 1\,\%$) for objects with $V \lesssim 11.5$\,mag.

SuperWASP collected data on \tyc{} and TZ\,Ari from 2004 to 2010, and from 2008 to 2011, respectively. The light curves were extracted and de-trended following the methods of \citet{Tamuz2005}, which were designed to remove instrumental systematics while preserving astrophysical signals.

\subsection{Las Cumbres Observatory (LCOGT)}

We obtained $V$ and $I'$-band images of \tyc{} using the 40\,cm telescopes of LCOGT \citep{Brown2013} at the Haleakal\=a{} (Maui) and Teide (Canary Islands) sites at 63 different epochs between June 28 and September 4, 2018. We acquired 50 individual exposures of 10\,s in the $V$-band and 5\,s in the $I'$-band per visit. The telescopes are equipped with a 3k\,$\times$\,2k SBIG CCD camera with a pixel scale of $0 \farcs 571$ providing a field of view of $29 \farcm 2 \times 19 \farcm 5$. The weather at the observatories was clear in general; the average seeing varied from $1 \farcs 3$ to $3 \farcs 1$. The raw data were processed using the {\tt BANZAI} pipeline \citep{McCully2018}, which includes bad pixel, bias, dark, and flat field corrections for each individual night. Aperture photometry was obtained for \tyc{} and several reference stars in the field using apertures of 11 pixels ($6 \farcs 3$). These apertures minimize the dispersion of the relative differential photometry between the target and reference stars.

TZ\,Ari was observed in the $I'$ band from the Siding Spring, Cerro Tololo, McDonald, and South African Astronomical Observatory sites of the LCOGT. A total of 538 photometric measurements were taken between September 29, 2019 and February 17, 2020. The data were reduced in the same way as those of \tyc.

\subsection{Sierra Nevada Observatory (OSN)}

Photometric observations of \tyc{} and TZ\,Ari were also collected at Observatorio de Sierra Nevada (OSN), Granada, Spain, using the T90 and T150
telescopes. T90 is a 90\,cm Ritchey-Chr\'etien telescope equipped with a VersArray 2k\,$\times$\,2k pixel CCD camera, providing a field of view of $13 \farcm 2 \times 13 \farcm 2$. The camera is based on a back-illuminated Marconi-EEV CCD42-4 with optimized response
in the ultraviolet \citep{Rodriguez2010}. Observations of \tyc{} were collected in Johnson $V$ and $R$ filters on 50 nights during the period between May and August 2018. Each CCD frame was corrected in a standard way for bias and flatfielding. All CCD measurements were obtained by the method of synthetic aperture photometry. Different aperture sizes were tested in order to choose the best one for our observations. A number of nearby and relatively bright stars within the frames were selected as reference stars. The data, in each filter, are presented as magnitude differences, normalized to zero. Outliers due to bad weather conditions or high airmass have been removed.

T150 is a 150\,cm Ritchey-Chr\'etien telescope equipped with a back-illuminated Andor Ikon-L DZ936N-BEX2-DD 2k\,$\times$\,2k CCD camera, with a resulting field of view of $7 \farcm 92 \times 7 \farcm 92$.
The camera is cooled thermo-electrically to $-100^{\circ}$C for negligible dark current. The observations of TZ\,Ari were collected in Johnson $V$ and $R$ filters on 42 nights during the period September 2019 to January 2020, and reduced in the same way as for T90.

\subsection{Montsec Observatory (OdM)}

Observations with the 80\,cm Joan Or\'o{} telescope at Montsec Observatory (Sant Esteve de la Sarga, Catalonia) were conducted using its main imaging camera LAIA, a 4k\,$\times$\,4k back-illuminated CCD with a pixel scale of $0\farcs4$ and a field of view of $30'$, using a Johnson $R$ filter. Several blocks of five images were obtained per night during the visibility periods of TZ\,Ari, resulting on 3115 measurements from February 2019 to March 2020. The images were calibrated with dark frames, bias, and flat fields, using the {\tt ICAT} pipeline \citep{Colome2006}. Differential photometry was extracted with {\tt AstroImageJ} \citep{Collins2017}, using the aperture size and the set of comparison stars that minimized the rms of the photometry.

\section{Spectroscopy}
\label{Spec}

\subsection{CARMENES}
\label{CARMENES}

\tyc{} and TZ\,Ari are included in the CARMENES radial-velocity survey of M dwarfs \citep{Quirrenbach2016, Reiners2018}, which is currently being conducted with the 3.5\,m telescope at the Calar Alto observatory (CAHA) in Almer\'ia, Spain. The CARMENES instrument was designed and optimized specifically for this survey. It consists of two cross-dispersed \'echelle spectrographs covering the wavelength regions 5\,200--9\,600\,\AA{} (VIS) and 9\,600--17\,100\,\AA{} (NIR) with a spectral resolution of $\mathcal{R} = 94\,600$ and $\mathcal{R} = 80\,400$, respectively \citep{Quirrenbach2014, Quirrenbach2018}.

We obtained 94 VIS and 96 NIR observations of \tyc,{} with exposure times ranging from 900 to 1800\,s, depending on seeing and sky transparency. TZ\,Ari was observed 93 times with the VIS and 89 times with the NIR spectrograph, using similar exposure times. All the spectra were accompanied by simultaneous Fabry-P\'erot spectra; the FP etalons were calibrated during daytime with Th-Ne, U-Ar, and U-Ne hollow-cathode lamps to establish an absolute wavelength scale.

All data were processed with the standard CARMENES pipelines \citep{Caballero2016b}. The data reduction pipeline {\tt caracal} applies standard corrections for detector bias and dark current, traces the \'echelle orders, extracts the spectra, and provides the wavelength calibration. The resulting fully reduced visible-light and near-infrared spectra are then analyzed with a second pipeline, {\tt serval} \citep{Zechmeister2018}. This process computes the RV of each observation by cross-correlating the corresponding spectrum with a reference template constructed from all observed spectra of the same star. In addition, {\tt serval} provides a number of spectral line indices\footnote{A line index $l$ can be converted into a pseudo-equivalent width of the line with the relation ${\rm pEW} = (1-l) \Delta \lambda$.}, which can be useful diagnostic tools for stellar variability, as well as indicators of the differential line width (dLW) and of the wavelength-dependence of the RV \citep[chromatic index, CRX; for details see][]{Zechmeister2018}. Further small corrections to the observed RVs are applied as described by \citet{Trifonov2018}.

The CARMENES RVs of \tyc{} and TZ\,Ari determined in this way are available from CDS. We note that for stars with spectral types of M1\,V (\tyc) and even as late as M5\,V (TZ\,Ari), the spectral information content in the VIS channel of CARMENES is much higher than in the NIR \citep{Reiners2018}. This leads to substantially smaller formal errors for the VIS data (independent of any calibration errors), as the VIS and NIR observations are executed simultaneously and thus with nearly identical integration times. Therefore, here we analyze mainly the VIS data, and we use the NIR data only to validate the planetary nature of TZ\,Ari\,b (see Sect.~\ref{act_cycle}).

\subsection{Archival data}
\label{Archival}

To complement our CARMENES data, we searched publicly available databases for precise RV time series of our two targets. No additional data were found for \tyc{}, but TZ\,Ari has been observed with HIRES/Keck and with HARPS.

The High Resolution Echelle Spectrograph \citep[HIRES,][]{Vogt1994}, installed at the right Nasmyth focus of the Keck\,I telescope, has been used extensively for RV monitoring with the iodine cell method. RV time series of 1624 stars obtained with this instrument were published by \citet{Butler2017}. We have used an improved version of these data \citep{Tal-Or2019}, which have been corrected for small nightly zero points and for a jump due to a CCD upgrade in 2004. This data set contains 54 observations of TZ\,Ari spread out over 13 years, with typical formal precision of 3.5\,m\,s$^{-1}$. \citet{Butler2017} also tabulated the Ca\,{\sc ii}\,H\&K $S$ index and an H$\alpha$ index, but the values of the $S$ index scatter widely (which is no surprise given the very low flux of M stars in the Ca\,{\sc ii}\,H\&K spectral region), and there are many invalid entries for the H$\alpha$ index. We therefore did not use any activity indicators from HIRES.

The HARPS instrument \citep{Mayor2003} at the European Southern Observatory's 3.6\,m telescope on La Silla, Chile, has pioneered the use of highly stabilized vacuum spectrographs for precise RV measurements. Twenty-five observations of TZ\,Ari with a typical RV precision of 4\,m\,s$^{-1}$ were found in the ESO archive. \citet{Trifonov2020b} have re-processed the archival HARPS data with {\tt serval}, yielding an improved RV precision for most stars. In the case of TZ\,Ari, however, we find the results from the original HARPS pipeline to be of superior quality; this is probably because the re-processing was not tuned for spectral types later than early M. We therefore used the HARPS RVs as retrieved from the ESO archive. In view of the relatively small number of HARPS data points, we did not attempt to extract time series of activity indicators from them.

\subsection{Periodograms}
\label{Periodograms}

\begin{figure}
   \centering
   \includegraphics[width=\hsize]{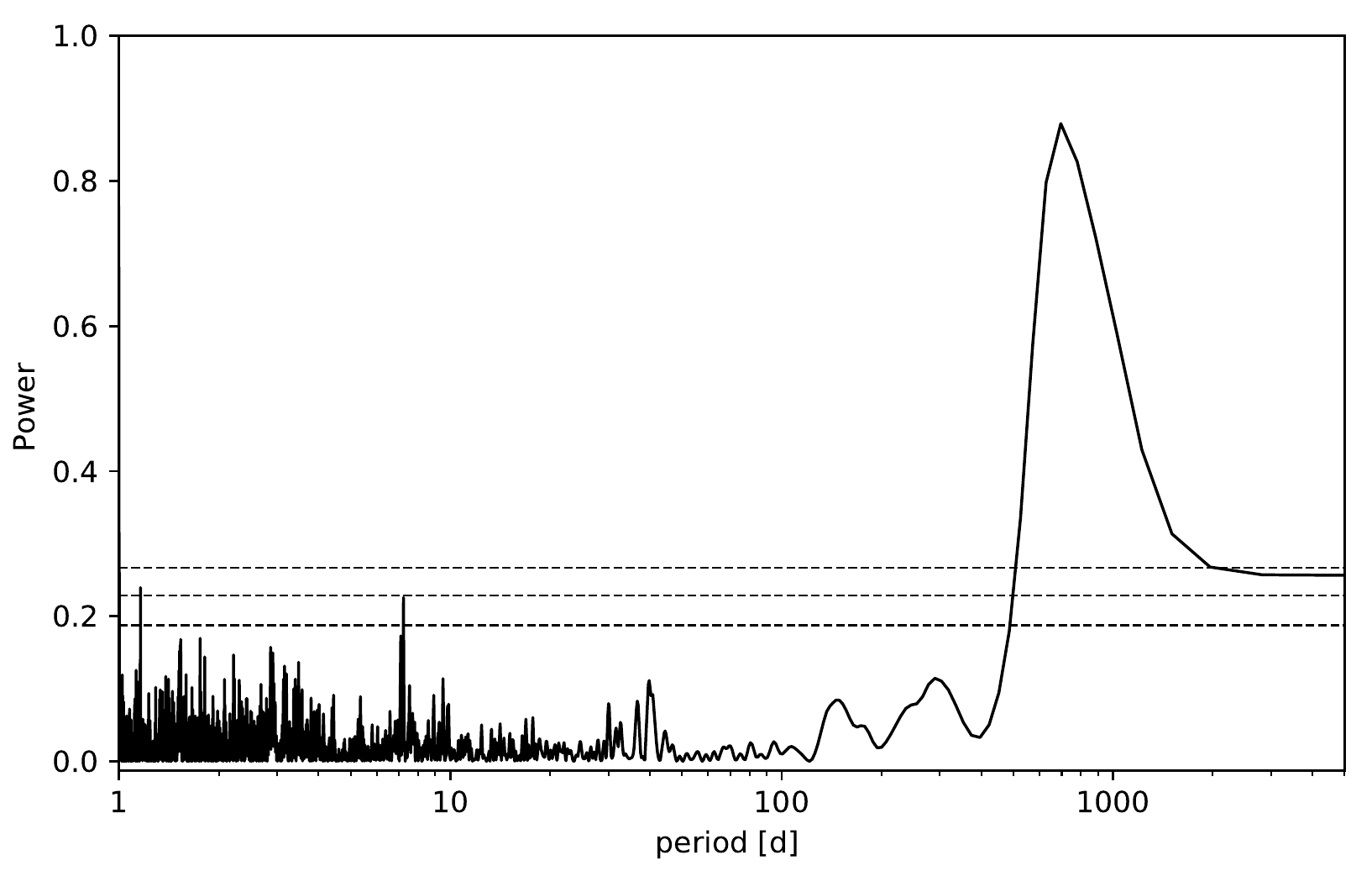}
      \caption{GLS periodogram of the RV time series of \tyc{} from CARMENES. The prominent peak at 696\,d can be attributed to a planet. The marginally significant peak at 7.15\,d and its alias at 1.15\,d are due to the window function of the sampling. The insignificant but noticeable power near 40\,d corresponds to the rotation period of the star. False-alarm probabilities of 0.1, 0.01, and 0.001 are shown as dashed horizontal lines.}
         \label{Fig_TYC_Periodogram}
   \end{figure}

\begin{figure}
   \centering
   \includegraphics[width=\hsize]{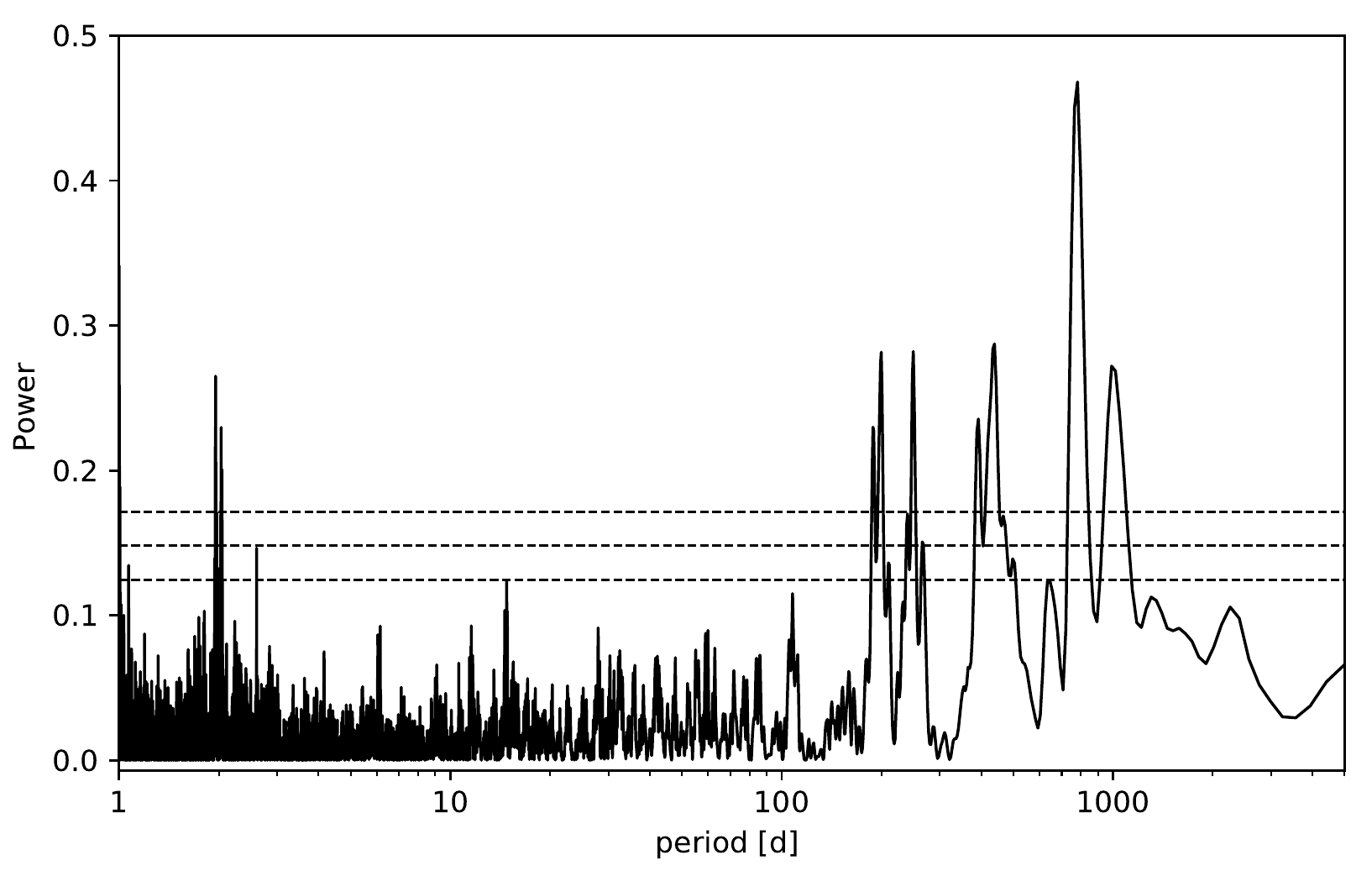}
      \caption{GLS periodogram of the combined RV time series of TZ\,Ari from CARMENES, HIRES, and HARPS. The prominent peaks at 781\,d and 1.96\,d correspond to a planet and to the stellar rotation period, respectively. False-alarm probabilities of 0.1, 0.01, and 0.001 are shown as dashed horizontal lines.}
         \label{Fig_TZ_Periodogram}
   \end{figure}

Figures~1 and 2 show generalized Lomb-Scargle \citep[GLS,][]{Zechmeister2009} periodograms of the RV time series of \tyc{} and TZ\,Ari, respectively. The periods apparent in these figures are analyzed in detail in the subsequent sections. Here, we give a brief overview. We attribute the most prominent peak in the \tyc{} periodogram at 696\,d to a planet, whereas the weaker peak at 7.15\,d and its one-day alias at 1.15\,d are due to the window function of the time series sampling. There is also some power near 40\,d, which corresponds to the rotation period of the star, as corroborated by the photometric data and spectroscopic indicators. The most significant peak in the GLS diagram of TZ\,Ari at 781\,d is also attributed to a planetary companion. The stellar rotation gives rise to the peak at 1.96\,d. The other peaks in Fig.~\ref{Fig_TZ_Periodogram} are related to these two through harmonics and their aliases; they will thus disappear in the residuals after modeling the planetary and rotational signals (see Fig.~\ref{Fig_TZ_Periodogram_Res}).

\section{Rotation periods and activity}
\label{Rot}

\subsection{Rotation period of \tyc}
\label{Rot_Tyc}

\begin{table}
\centering
\small
\caption{Rotation periods of \tyc{} and TZ\,Ari estimated from different data sets.} \label{tab:rotation}
\begin{tabular}{lcc}
\hline\hline
\noalign{\smallskip}
 & \tyc & TZ\,Ari \\
Data set & Period [d] & Period [d] \\
\hline
\noalign{\smallskip}
\multicolumn{3}{c}{\em Photometry}\\
\noalign{\smallskip}
SuperWASP & 41.0 & 1.96 \\
LCOGT $V$ & 37.7 & -- \\
LCOGT $I'$ & 42.4 & -- \\
OSN $R$ & 38.9 & 1.94 \\
OSN $V$ & 38.9 & 1.95 \\
Montsec $R$ & -- & (1.96) \\
\noalign{\smallskip}
\multicolumn{3}{c}{\em Spectroscopy}\\
\noalign{\smallskip}
RV & 39.8 & 1.96 \\
pEW(H$\alpha$) & 39.9 & 1.94 \\
Ca\,{\sc ii} IRT & 39.8 & 1.95 \\
CRX & -- & 1.96 \\
dLW & 39.1 & 1.96\\
\noalign{\smallskip}
\multicolumn{3}{c}{\em Adopted}\\
\noalign{\smallskip}
All & 40 $\pm$ 1 & 1.96 $\pm$ 0.02\\
\noalign{\smallskip}
\hline
\end{tabular}
\end{table}

The photometric light curves of \tyc{} from SuperWASP, LCOGT, and OSN all show a periodicity near 40\,d (see Table~\ref{tab:rotation}), indicating that this is the rotation period of the star. A subset of the SuperWASP data was already analyzed by \citet{Diez_Alonso2019}, who arrived at a value of $41.0 \pm 1.7$\,d. The CARMENES pipeline delivers several indicators that can also be used to determine rotation periods \citep{Fuhrmeister2019,Lafarga2021}; the most useful among these are the pEW of the H$\alpha$ line, an index of the Ca\,{\sc ii} infrared triplet (IRT), the CRX, and the dLW. Three of these (H$\alpha$, IRT, and dLW) show the 40\,d periodicity very clearly (see Appendix~\ref{Indicators} for two examples), and there is also a strong peak near this value in the periodogram of the RVs. Taking into account the dispersion between these individual indicators, we conclude that the rotation period of \tyc{} is $40 \pm 1$\,d.

\subsection{Activity of \tyc}
\label{Act_Tyc}

\begin{figure}
   \centering
   \includegraphics[width=\hsize]{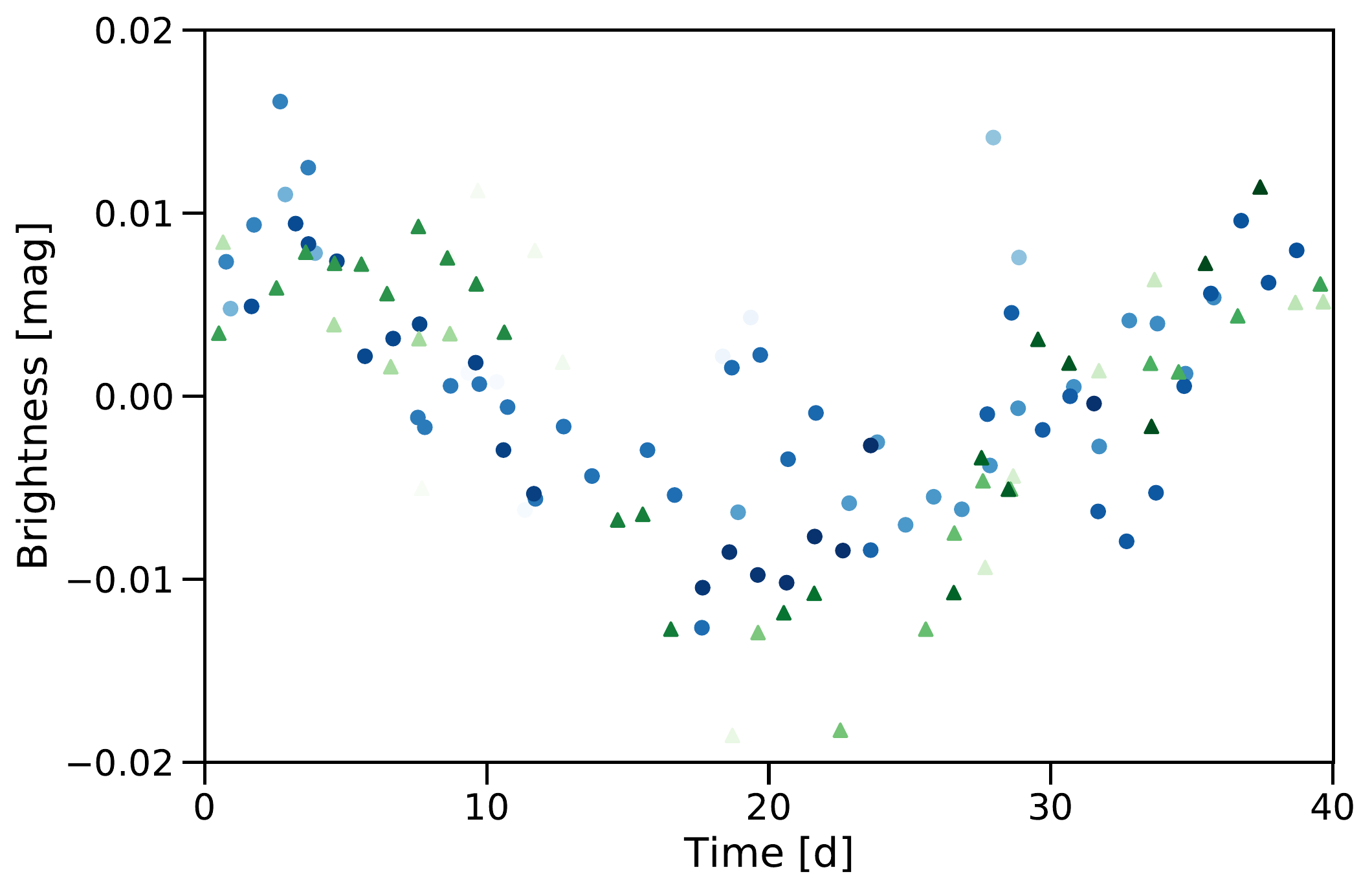}
      \caption{Differential photometry of \tyc, phase-folded to $P = 40.0203$\,d. Each data point represents measurements averaged over one night. SuperWASP data are shown as blue circles, and OSN $R$-band data are represented by green triangles. Within each of the two time series, the symbols change from light to dark with progressing time. See the text for more details, }
         \label{Fig_TYC_Photom}
   \end{figure}

\begin{figure}
   \centering
   \includegraphics[width=\hsize]{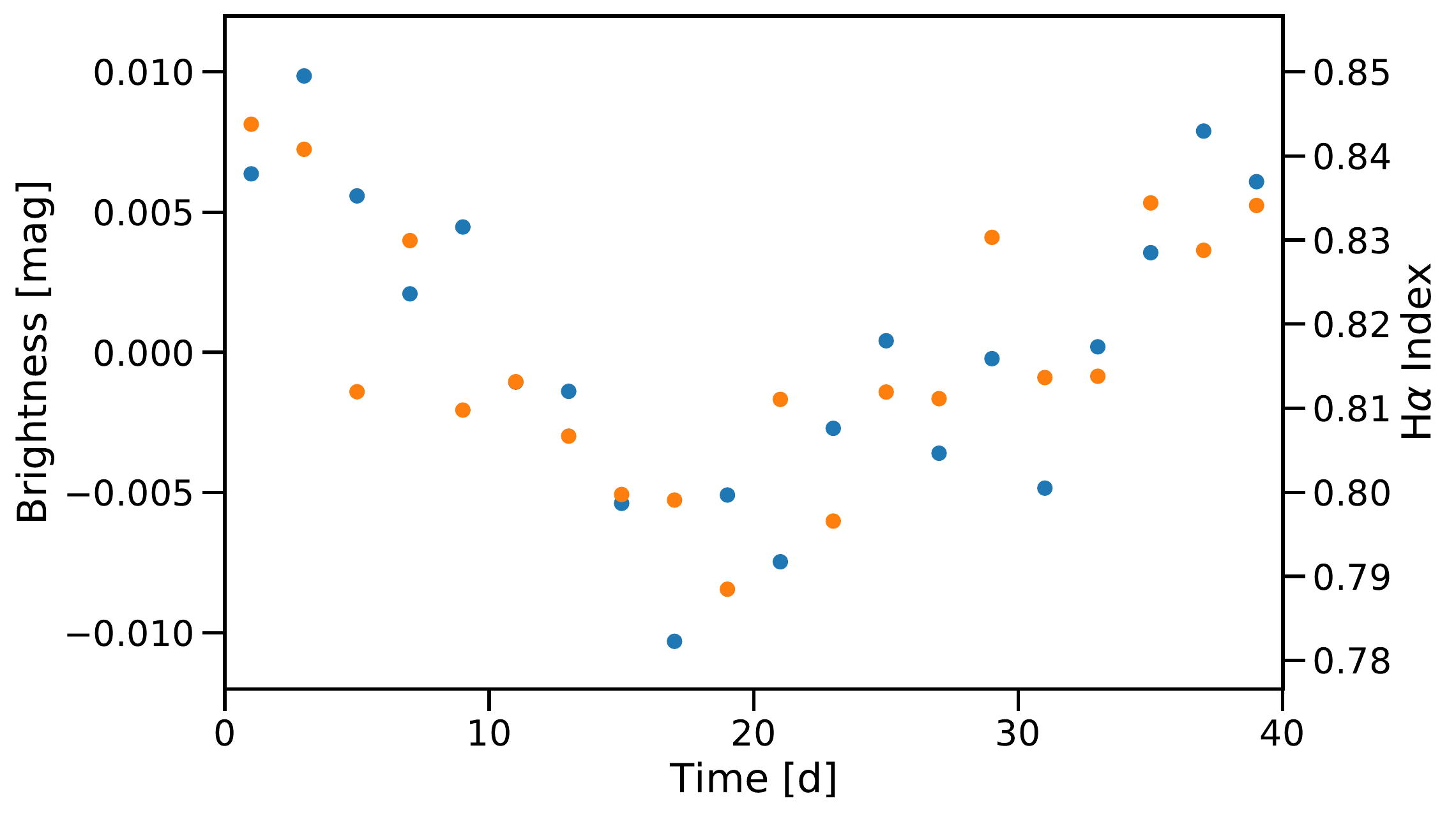}
      \caption{Photometry and H$\alpha$ index of \tyc, phase-folded to $P = 40.0203$\,d and averaged into 20 equally spaced phase bins. The photometric data from SuperWASP and OSN ($R$-band) are shown in blue, and the H$\alpha$ line index from CARMENES is shown in orange.}
         \label{Fig_TYC_P_H}
   \end{figure}

A comparison between the time series of \tyc{} taken from 2004 to 2010 with SuperWASP and the more recent observations, for instance,\ with OSN in 2018, reveals a striking persistence and consistency of the photometric variability of the star, both in amplitude and phase. A combined periodogram of the SuperWASP and OSN data shows multiple narrow peaks with almost identical amplitudes near 40\,d, which are aliases of each other, resulting from the wide gap between the two data sets. We picked one of them at 40.0203\,d and phase-folded the SuperWASP and OSN data to this period, without any attempt to account for the differences between the filters used by the two instruments. Nonetheless, the two light curves match almost perfectly (see Fig.~\ref{Fig_TYC_Photom}), indicating that the photometric behavior of the star has remained remarkably stable for one and a half decades. A similar stable pattern was reported for the fully convective M4 dwarf V374\,Peg \citep{Vida2016}. This is to be contrasted with LQ\,Hya, a BY\,Dra-type variable of spectral type K2\,V, in which active areas on the stellar surface remain stable only on time scales less than a year \citep{Lehtinen2012}. Such observations provide valuable input for stellar dynamo models. A better understanding of the persistence of activity patterns is also important in the context of exoplanet detection, as stable RV modulations can easily be mistaken for the signature of a planet if the stellar rotation period is not known. Systematic studies of larger samples are therefore urgently needed, but the precision of current ground-based surveys such as ASAS is marginal for this type of work \citep{Suarez_Mascareno2016}.

The H$\alpha$ absorption line is slightly shallower in \tyc{} than in an inactive reference star \citep{Schoefer2019}. This means that the line core is partially filled by an emission component, qualifying \tyc{} as ``H$\alpha$ active.'' Figure~\ref{Fig_TYC_P_H} shows the same photometric data as Fig.~\ref{Fig_TYC_Photom}, in 20 equally spaced phase bins, along with the H$\alpha$ line index as defined by \citet{Zechmeister2018} from the CARMENES spectra. From this plot it is evident that brightness and line index track each other very closely and consistently. Whenever the star is fainter (more positive magnitude), the line index is larger, namely, the core of the H$\alpha$ line is more filled-in. This suggests that dark spotted patches on the stellar surface are accompanied by H$\alpha$ emission regions.

\subsection{Rotation period of TZ\,Ari}
\label{Rot_TZ}

The $R$-band photometric variability amplitude of TZ\,Ari is only $\sim 3$\,mmag, which makes it difficult to unambiguously identify its rotation period from ground-based monitoring data. Since TZ\,Ari will not be observed by {\em TESS} during cycles 1 through 4 (Sectors 1 through 55), we therefore start with the CARMENES spectroscopy. The RVs show a strong modulation with a period of 1.96\,d and an amplitude that is not completely stable, but instead varying around a typical level of $\sim 10$\,m\,s$^{-1}$ (see also Figs.~\ref{Fig_TZ_GP_RV_Zoom} and \ref{Fig_TZ_GP_RV_Res}). These RV variations are chromatic, and thus cannot be caused by Doppler shifts due to an orbiting companion. Variations with the same periodicity are also observed in the H$\alpha$ and Ca\,{\sc ii} IRT lines (see Table~\ref{tab:rotation} and Appendix~\ref{Indicators}), strengthening the conclusion that 1.96\,d is the rotation period.

With this information in mind, the photometry can now be interpreted with confidence. The periodograms of the OSN $R$ and $V$ data both show a strong peak at 1.96\,d, along with its slightly weaker daily alias at 2.03\,d. The strongest peak for SuperWASP, at 131\,d, can be attributed to instrumental effects, but again there are highly significant peaks at 1.96\,d and 2.03\,d. A peak at 1.96\,d is also found for the Montsec time series, although it is not the highest in that periodogram. The LCOGT photometry is not precise enough to be useful for the determination of the rotation period of TZ\,Ari. Overall, the photometric variations thus fully confirm that the rotation period is $1.96 \pm 0.02$\,d.

\subsection{Activity of TZ\,Ari}
\label{Act_TZ}

The H$\alpha$ line in TZ\,Ari is observed strongly in emission with temporally variable equivalent width, qualifying TZ\,Ari also as an H$\alpha$ active star \citep{Schoefer2019}. White-light flares with amplitude up to $\Delta V = 0.87$ were found in the light curve of TZ\,Ari from the ASAS-SN monitoring survey \citep{Rodriguez_Martinez2020}. These findings are to be contrasted with the small rotational modulation of the stellar brightness. We can thus plausibly infer that chromospheric activity occurs rather uniformly over the stellar surface, or in regions with a uniform longitudinal distribution. However, the strong rotation-related RV variations point towards a significantly structured photosphere.

\section{Modeling of the radial velocities and planet parameters}
\label{Modeling}

\subsection{Gaussian process modeling}
\label{GPModeling}

Stellar rotation induces quasi-periodic variations in the brightness, activity indicators, line shapes, and RV as spots, plage, and active regions rotate into and out of view \citep{Dumusque2014}. These variations can be used to determine the rotation period and to characterize various aspects of stellar activity, as discussed in the previous section. On the other hand, rotational modulation of the RVs may obscure the Keplerian signature of exoplanets, mimic spurious planets, or affect the planet parameters derived from models of the RV data. Therefore, any realistic comprehensive model of the RVs should be the sum of two components describing Keplerian orbits and stellar variability, respectively. The latter can be represented by a Gaussian process \citep[e.g.][]{Haywood2014}, and the best-fit model is found by maximizing the likelihood, varying the Keplerian parameters and the hyper-parameters of the GP simultaneously. Several software packages are available to perform this task; we use the {\tt Exo-Striker} \citep{Trifonov2019}.

An important aspect of the RV modeling is the choice of the kernel function describing the noise covariance in the GP model. A physically well-motivated choice is the quasi-periodic kernel \citep[e.g.,][]{Rajpaul2015}:
\begin{equation}\label{QPkernel}
  k(\tau) = a \exp \left[ - \frac{\tau^2}{2\,l^2} - \Gamma \sin^2 \left(\frac{\pi\,\tau}{P}\right) \right]~~,
\end{equation}
where the first term in the square bracket describes a stellar surface structure decaying with a lifetime, $l$, and the second term the periodic rotation of this structure into and out of view. For computational efficiency reasons, this kernel is sometimes approximated by \citep{Foreman-Mackey2017}
\begin{equation}\label{Rotkernel}
  k(\tau) = \frac{a}{2+b}\,e^{-c\,\tau}\left[ \cos\left(\frac{2\,\pi\,\tau}{P}\right) + (1+b) \right]~~,
\end{equation}
which has similar properties. We tried to model the RV time series of \tyc{} and TZ\,Ari with this kernel, but found it problematic. The issue is that the planets orbiting these stars have periods that are much longer than the respective rotational periods and the time scales on which the surface structure may evolve. Models described by kernels such as those given by Eqs.~\ref{QPkernel} and \ref{Rotkernel} are not constrained to have zero mean on long time scales, as the noise covariance matrix vanishes for $\tau \gg l$, or $\tau \gg 1/c$, respectively. This means that the GP tends to ``absorb'' part of the planetary signal, artificially reducing the inferred value of the semi-amplitude, $K$, as well as the computed significance \citep{Morales2019}.

Since, in the present context, we are mostly interested in the planet parameters and not so much in a faithful representation of the stellar RV variation through the GP model, we adopted a simple harmonic oscillator (SHO) as the kernel function; a similar model was also used by \citet{Ribas2018}. With the abbreviation
\begin{equation}\label{eta_def}
  \eta = \sqrt{1-\frac{1}{4Q^2}}~~,
\end{equation}
the SHO kernel is given by \citep{Foreman-Mackey2017}:
\begin{equation}\label{SHOkernel}
k(\tau) = S \omega_0 Q \exp \left(- \frac{\omega_0 \tau}{2 Q} \right) \left[ \cos (\eta \omega_0 \tau) + \frac{1}{2 \eta Q } \sin (\eta \omega_0 \tau) \right]~~.
\end{equation}
Here $S$ is the amplitude of the power spectral density of the process, $\omega_0$ the angular frequency, and $Q$ the quality factor of the oscillator. While it is possible to relate $\omega_0 = 2 \pi / P$ and $Q = l \omega_0 /2$ to the rotation period, $P,$ and typical lifetime, $l,$ of the stellar structure, we emphasize that
the model was chosen to model the stellar RV variations without affecting the long-period Keplerian signals -- and not specifically to provide a physical description of the former.

\subsection{\tyc}

\subsubsection{Single-planet Keplerian fit}

    \begin{table}[ht]
    \centering
    \caption{Single-planet Keplerian and Gaussian process model fit to the RV data of \tyc{} from CARMENES VIS. The corresponding priors for the GP are listed in Table~\ref{tab:TYC_GP_Priors}.}
    \label{tab:TYC_RVfits}

    \begin{tabular}{lrr}

    \hline\hline
    \noalign{\vskip 0.7mm}            & \multicolumn{2}{c}{\tyc{}\,b}\\
    Parameter \hspace{0.0 mm}         &  Keplerian                     & GP model\\
    \hline \noalign{\vskip 0.7mm}

        $K$ [m\,s$^{-1}$]             &        12.28$_{-0.56}^{+0.56}$ & 12.02$_{-0.46}^{+0.47}$ \\ \noalign{\vskip 0.9mm}
        $P$ [d]                       &     694.86$_{-10.75}^{+11.22}$ &    691.90$_{-8.61}^{+8.77}$ \\ \noalign{\vskip 0.9mm}
        $e$                           &       0.05$_{-0.03}^{+0.05}$   &      0.05$_{-0.03}^{+0.04}$ \\ \noalign{\vskip 0.9mm}
        $\omega$ [deg]                &      15.10$_{-63.62}^{+77.88}$ &      3.71$_{-46.94}^{+64.06}$ \\ \noalign{\vskip 0.9mm}
        $M_{\rm 0}$ [deg]             &     269.35$_{-76.61}^{+62.35}$ &    278.98$_{-62.75}^{+46.18}$ \\ \noalign{\vskip 0.9mm}
        $a$ [au]                      &       1.22$_{-0.02}^{+0.02}$   &      1.22$_{-0.02}^{+0.02}$ \\ \noalign{\vskip 0.9mm}
        $m \sin i$ [$M_{\rm Jup}$]    &       0.34$_{-0.02}^{+0.02}$   &      0.33$_{-0.02}^{+0.02}$ \\ \noalign{\vskip 0.9mm}
        RV$_{\rm off}$ [m\,s$^{-1}$]  &      $-0.01_{-0.42}^{+0.42}$   &     $-0.23_{-0.35}^{+0.34}$ \\ \noalign{\vskip 0.9mm}
        RV$_{\rm jit}$ [m\,s$^{-1}$]  &       3.21$_{-0.29}^{+0.33}$   &      2.06$_{-0.29}^{+0.32}$ \\ \noalign{\vskip 0.9mm}
        GP$_{\rm SHO}$ $S$ [m$^{2}$\,s$^{-2}$\,d] &  \multicolumn{1}{c}{--}        &      1.72$_{-0.72}^{+0.88}$ \\ \noalign{\vskip 0.9mm}
        GP$_{\rm SHO}$ $Q$            &  \multicolumn{1}{c}{--}        &     23.61$_{-10.27}^{+23.43}$ \\ \noalign{\vskip 0.9mm}
        GP$_{\rm SHO}$ $\omega_0$ [d$^{-1}$] &  \multicolumn{1}{c}{--}        &      0.160$_{-0.003}^{+0.004}$ \\ \noalign{\vskip 0.9mm}
        $rms$ [m\,s$^{-1}$]           &       3.41                     &      2.33 \\
        $\Delta$BIC                   & \multicolumn{2}{c}{$-32$} \\
        $N_{\rm RV data}$             & \multicolumn{2}{c}{94} \\
        Epoch                         & \multicolumn{2}{c}{2457558.647} \\

        \hline \noalign{\vskip 0.7mm}

    \end{tabular}
    \end{table}

\begin{figure}
   \centering
   \includegraphics[width=\hsize]{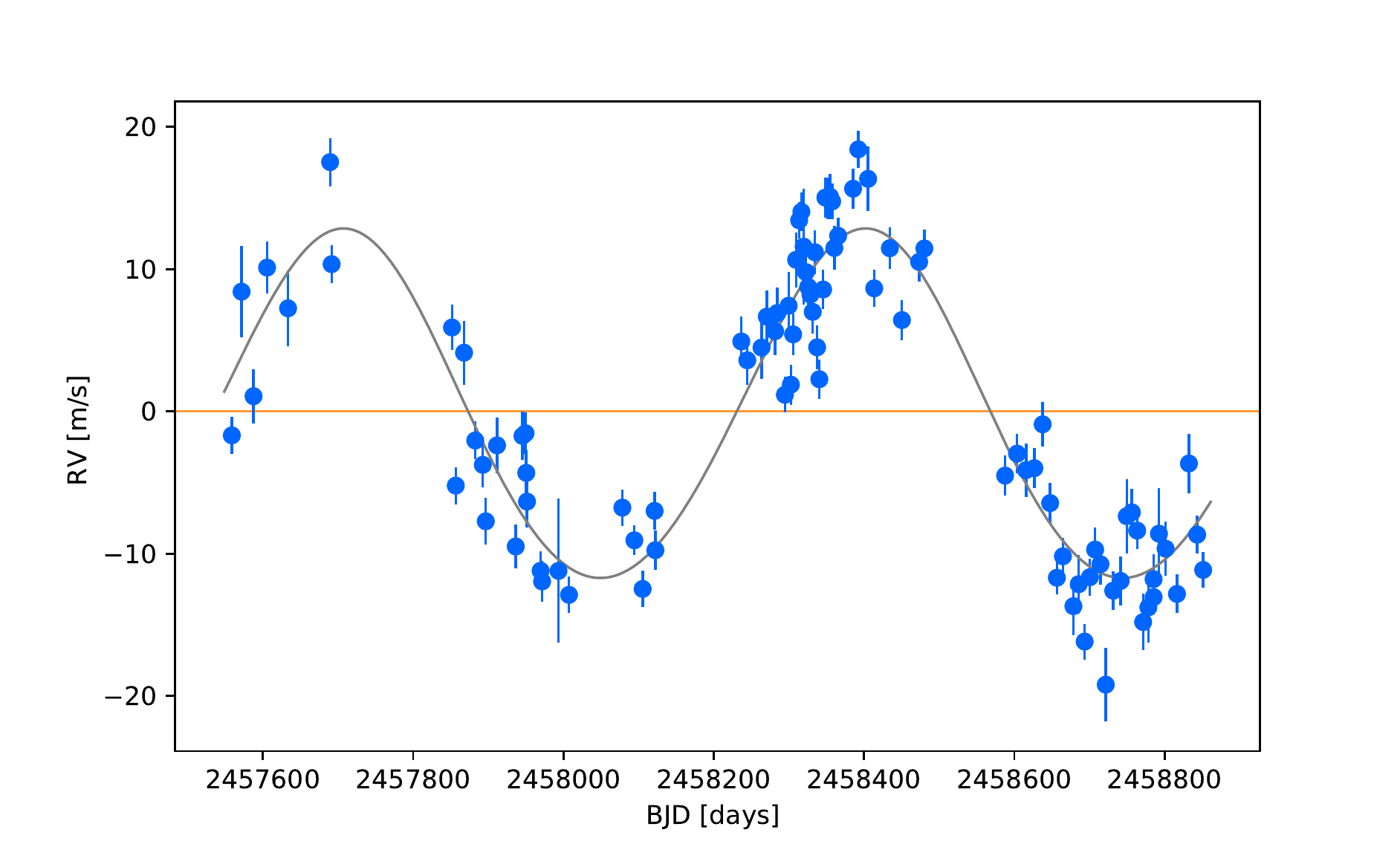}
      \caption{Single-planet Keplerian fit to the RV data of \tyc{} from CARMENES VIS. The orbital parameters are listed in the left column of Table~\ref{tab:TYC_RVfits}.}
         \label{Fig_TYC_Single_RV}
   \end{figure}

Before proceeding with the full GP modeling, we first fit a simple model consisting of seven free parameters -- the five standard orbital elements: velocity semi-amplitude, $K$, period, $P$, eccentricity, $e$, argument of periastron, $\omega$, and mean anomaly at the reference epoch, $M_0$, as well as a radial-velocity zero point and a Gaussian white noise RV ``jitter'' term -- to the CARMENES RV data of \tyc{}. We
explored the parameter space with a Markov chain Monte Carlo (MCMC) procedure to determine the $1 \sigma$ confidence intervals of the best-fit parameters.
The results are tabulated in the left column of Table~\ref{tab:TYC_RVfits} and shown in Fig.~\ref{Fig_TYC_Single_RV}, plotted together with the RV data. A planet with $P = 695$\,d is evidently detected at high significance ($\sim 20\,\sigma$); we give it the designation \tyc\,b. Because of its simplicity, the plain single-planet model provides a useful reference point and lends confidence to the results of the more sophisticated Gaussian process modeling described in the subsequent section.

We double-checked that no signal near 695\,d is present in the photometric data or in the spectroscopic time series other than the RV (i.e., in line indices, dLW, or CRX). While these tests do not conclusively rule out a stellar origin of the RV signal, they render a planetary nature much more likely, as explained in more detail in Sect.~\ref{act_cycle}.

\subsubsection{Gaussian process model}
\label{TYC_GP_model}

\begin{figure}
   \centering
   \includegraphics[width=\hsize]{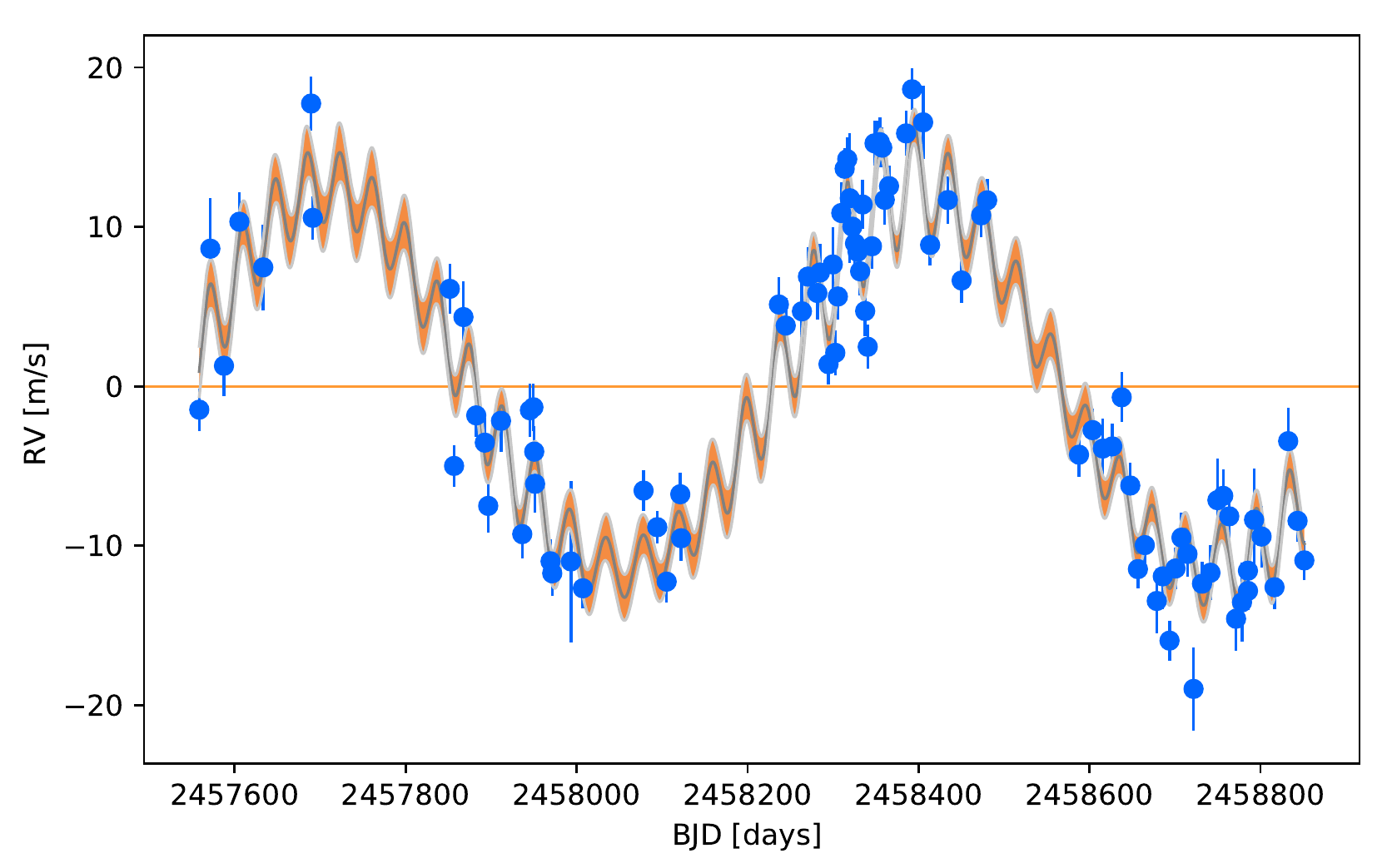}
      \caption{Gaussian process model fit to the RV data of \tyc{} from CARMENES VIS. The orbital parameters are listed in the right column of Table~\ref{tab:TYC_RVfits}. Here and in the following figures the GP model and its $1 \sigma$ credibility range are shown in grey and brown, respectively.}
         \label{Fig_TYC_GP_RV}
   \end{figure}

\begin{figure}
   \centering
   \includegraphics[width=\hsize]{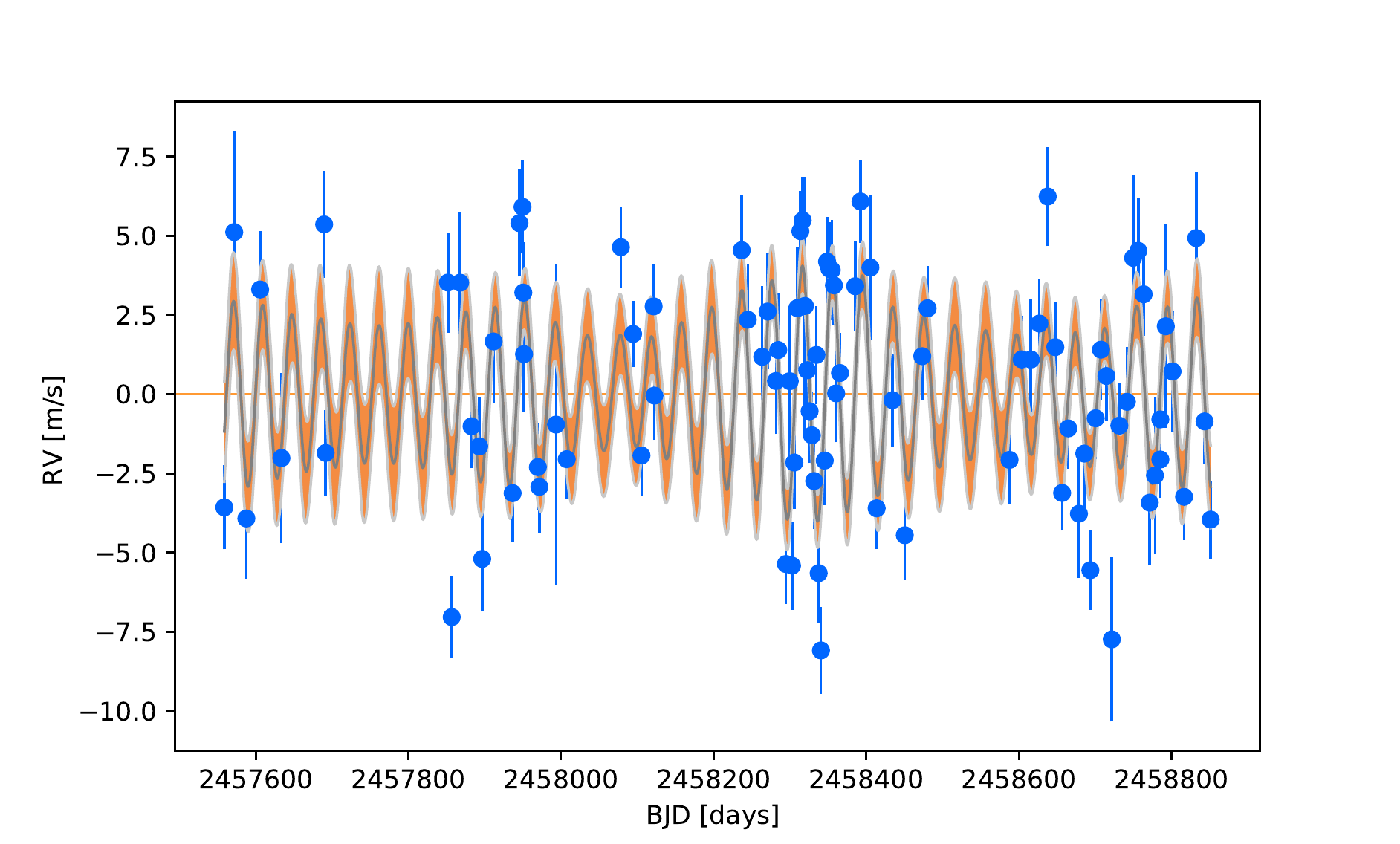}
      \caption{Gaussian process model fit to the RV data of \tyc{} from CARMENES VIS, as in Fig.~\ref{Fig_TYC_GP_RV}, but with the planet orbit fit subtracted. This figure shows that the SHO model reproduces the $\sim 40$-day quasi-periodic variations very well, without introducing any power on longer time scales, which could affect the best-fit planet parameters.}
         \label{Fig_TYC_GP_RV_Res}
   \end{figure}

\begin{figure}
   \centering
   \includegraphics[width=\hsize]{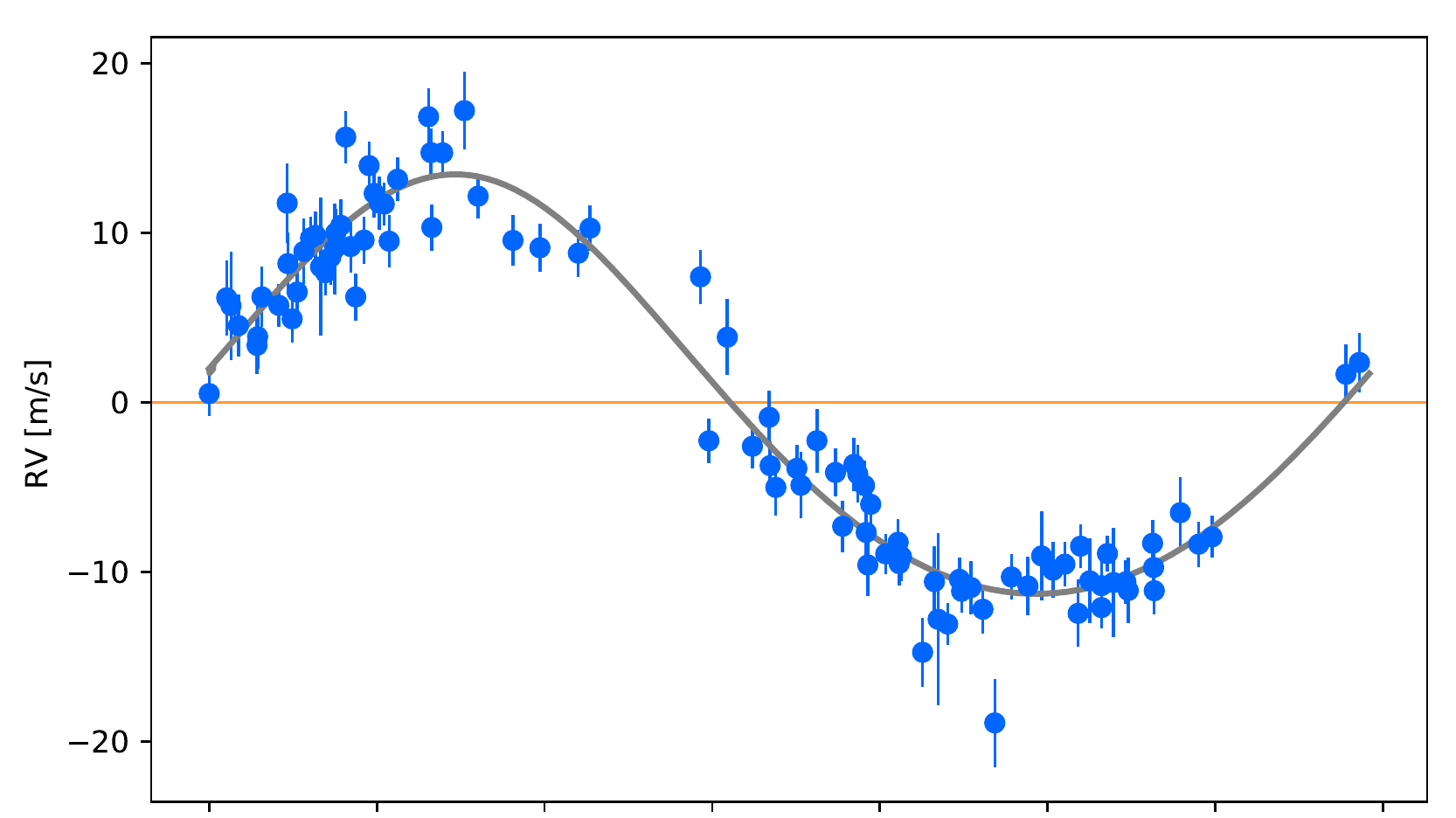}
   \includegraphics[width=\hsize]{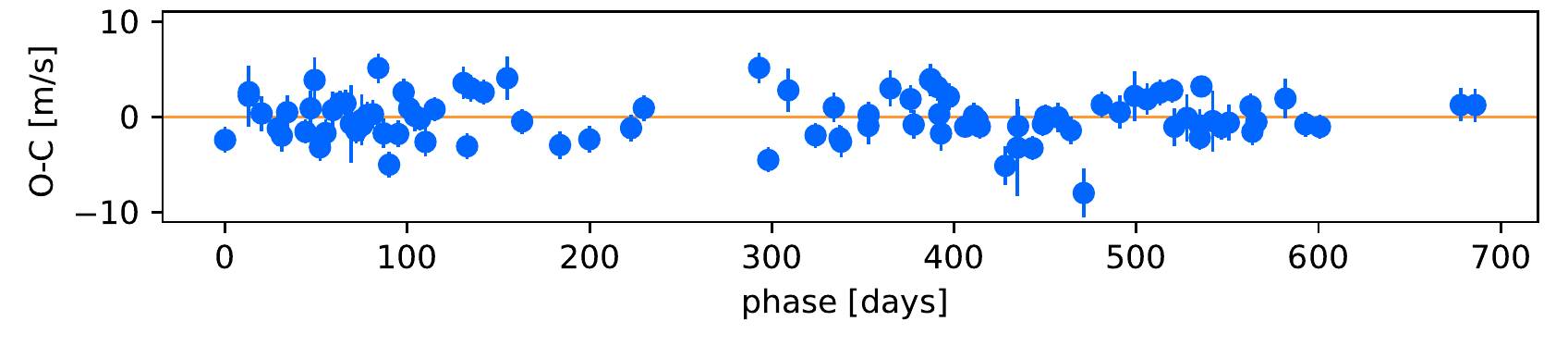}
         \caption{Gaussian process model fit to the RV data of \tyc{} from CARMENES VIS, as in Fig.~\ref{Fig_TYC_GP_RV}, but with the Gaussian process model of the stellar variability subtracted and folded to the period of \tyc{}\,b (691.90\,d).}
         \label{Fig_TYC_GP_RV_Folded}
   \end{figure}

\begin{figure}
   \centering
   \includegraphics[width=\hsize]{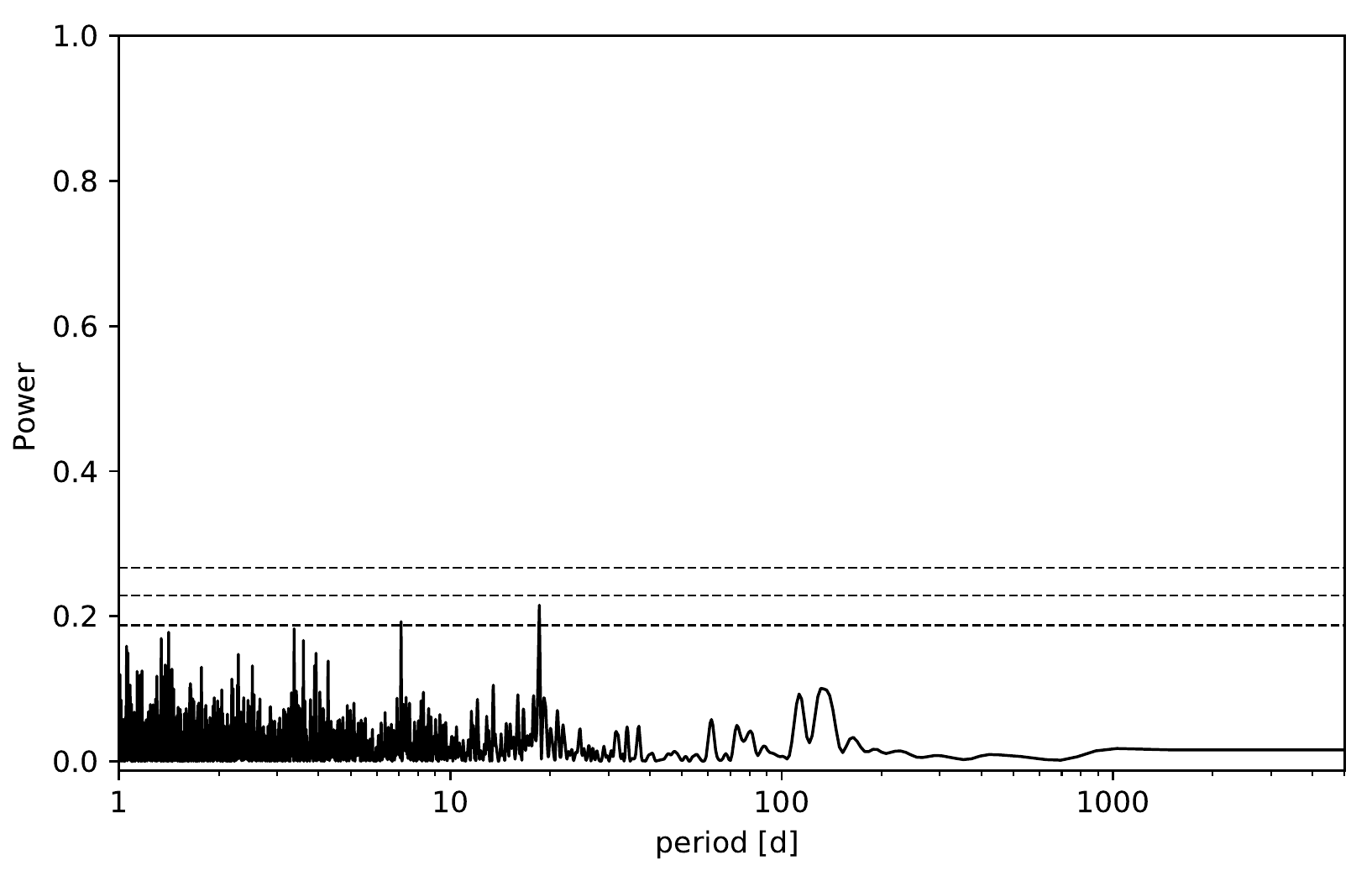}
      \caption{GLS periodogram of the residuals of the GP model for \tyc{}. False-alarm probabilities of 0.1, 0.01, and 0.001 are shown as dashed horizontal lines. The scales of the axes are identical to Fig.~\ref{Fig_TYC_Periodogram} for better comparison.}
         \label{Fig_TYC_Periodogram_Res}
   \end{figure}

To obtain a consistent picture of the RV variations of \tyc, we used a model that describes the planet with a Keplerian, and the rotational modulation with a Gaussian process, as explained in Sect.~\ref{GPModeling}. We performed an MCMC optimization of the Keplerian parameters and GP hyper-parameters with uninformative priors within bounds based on the single-planet fit, as tabulated in Appendix~\ref{GP_priors}. We employed 500 walkers and let them run for 700 steps each, discarding the first 200 samples as a burn-in phase. We checked that the process converged, and used a total of 81,354 accepted samples to compute distributions of the model parameters. The results are listed in the right column of Table~\ref{tab:TYC_RVfits}; more detailed information about the underlying distributions of the MCMC samples is shown in Appendix~\ref{cornerplots}. A comparison of both columns in Table~\ref{tab:TYC_RVfits} shows that all planet parameters are fully consistent with those obtained from the single-planet model, but the GP model is favored strongly as indicated by the difference in the Bayesian information criterion (BIC) of $-32$. The RV curve of the combined model is shown in Fig.~\ref{Fig_TYC_GP_RV}; both the planet and the stellar rotation are evidently reproduced well. This is further demonstrated in Figs.~\ref{Fig_TYC_GP_RV_Res} and \ref{Fig_TYC_GP_RV_Folded}, which show the contributions of the GP and of the Keplerian to the observed RV separately. We note that the short-term variations represented by the GP maintain a zero mean on longer time scales as intended; thus, the GP does not add or subtract power from the planetary signal. We therefore adopt the best-fit parameters of the GP model listed in the right-hand column of Table~\ref{tab:TYC_RVfits} for the planet \tyc\,b.

Figure~\ref{Fig_TYC_Periodogram_Res} shows that no significant power remains at any frequency in the residuals of the GP model. The only peak with a false-alarm probability of only a few per cent is at 18.6\,d, very close to the second harmonic of the rotation. We conclude that all our data on \tyc{} are represented well by the ``one planet plus rotational modulation'' model.

\subsection{TZ\,Ari}

\subsubsection{Single-planet Keplerian fit}

    \begin{table}
    \centering
    \caption{Single-planet Keplerian and Gaussian process model fit to the combined RV data of TZ\,Ari from HIRES, CARMENES VIS, and HARPS (labeled as 1, 2, and 3, respectively). The corresponding priors for the GP are listed in Table~\ref{tab:TZ_GP_Priors}.}
    \label{tab:TZ_RVfits}

    \begin{tabular}{lrr}     

    \hline\hline
\noalign{\vskip 0.7mm}            & \multicolumn{2}{c}{TZ\,Ari\,b}\\
    Parameter \hspace{0.0 mm}         &  Keplerian                     & GP model\\
    \hline \noalign{\vskip 0.7mm}

        $K$ [m\,s$^{-1}$]             &      21.11$_{-1.81}^{+1.91}$ &     18.84$_{-1.16}^{+1.30}$ \\ \noalign{\vskip 0.9mm}
        $P$ [d]                       &     772.05$_{-1.84}^{+2.41}$ &    771.36$_{-1.23}^{+1.34}$ \\ \noalign{\vskip 0.9mm}
        $e$                           &       0.49$_{-0.07}^{+0.06}$ &      0.46$_{-0.04}^{+0.04}$ \\ \noalign{\vskip 0.9mm}
        $\omega$ [deg]                &     325.67$_{-9.11}^{+8.70}$ &    321.79$_{-5.67}^{+5.32}$ \\ \noalign{\vskip 0.9mm}
        $M_{\rm 0}$ [deg]             &     267.96$_{-6.35}^{+8.20}$ &    268.04$_{-5.11}^{+5.73}$ \\ \noalign{\vskip 0.9mm}
        $a$ [au]                      &       0.88$_{-0.02}^{+0.02}$ &      0.88$_{-0.02}^{+0.02}$ \\ \noalign{\vskip 0.9mm}
        $m \sin i$ [$M_{\rm Jup}$]    &       0.23$_{-0.02}^{+0.02}$ &      0.21$_{-0.02}^{+0.02}$ \\ \noalign{\vskip 0.9mm}
        RV$_{\rm off}$ 1 [m\,s$^{-1}$]&      $-3.03_{-1.81}^{+1.81}$ &     $-3.96_{-1.44}^{+1.45}$ \\ \noalign{\vskip 0.9mm}
        RV$_{\rm off}$ 2 [m\,s$^{-1}$]&      $-0.93_{-1.16}^{+1.16}$ &     $-3.06_{-0.54}^{+0.54}$ \\ \noalign{\vskip 0.9mm}
        RV$_{\rm off}$ 3 [m\,s$^{-1}$]&      $-11.10_{-2.29}^{+2.46}$ &     $-8.59_{-1.94}^{+1.87}$ \\ \noalign{\vskip 0.9mm}
        RV$_{\rm jit}$ 1 [m\,s$^{-1}$]&      12.47$_{-1.29}^{+1.56}$ &      8.53$_{-1.38}^{+1.57}$ \\ \noalign{\vskip 0.9mm}
        RV$_{\rm jit}$ 2 [m\,s$^{-1}$]&       9.75$_{-0.74}^{+0.81}$ &      3.09$_{-0.43}^{+0.47}$ \\ \noalign{\vskip 0.9mm}
        RV$_{\rm jit}$ 3 [m\,s$^{-1}$]&       6.65$_{-1.56}^{+1.90}$ &      2.58$_{-1.55}^{+1.88}$ \\ \noalign{\vskip 0.9mm}
        GP$_{\rm SHO}$ $S$ [m$^{2}$\,s$^{-2}$\,d] & \multicolumn{1}{c}{--} & 0.022$_{-0.004}^{+0.004}$ \\ \noalign{\vskip 0.9mm}
        GP$_{\rm SHO}$ $Q$            &  \multicolumn{1}{c}{--}      &   625.64$_{-181.90}^{+208.03}$ \\ \noalign{\vskip 0.9mm}
        GP$_{\rm SHO}$ $\omega_0$ [d$^{-1}$] &  \multicolumn{1}{c}{--} &    3.214$_{-0.002}^{+0.002}$ \\ \noalign{\vskip 0.9mm}
        $rms$ [m\,s$^{-1}$]           &      10.37 &      5.38 \\
        $\Delta$BIC                   & \multicolumn{2}{c}{$-152$} \\
        $N_{\rm RV data}$             &        \multicolumn{2}{c}{172} \\
        Epoch                         & \multicolumn{2}{c}{2451411.05} \\
        \hline \noalign{\vskip 0.7mm}

    \end{tabular}



    \end{table}

\begin{figure}
   \centering
   \includegraphics[width=\hsize]{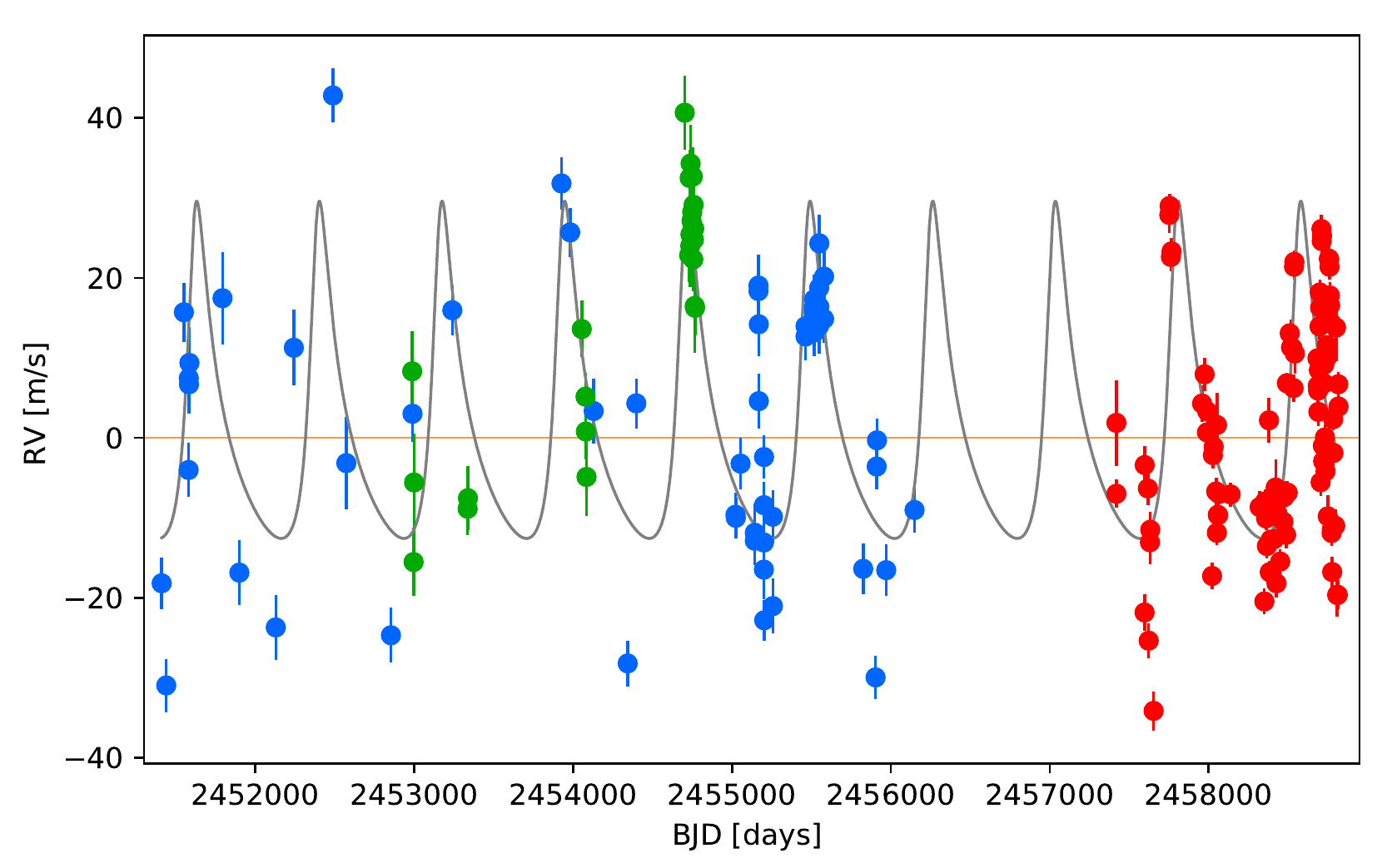}
      \caption{Single-planet Keplerian fit to the RV data of TZ\,Ari from HIRES (blue), HARPS (green), and CARMENES VIS (red). The orbital parameters are listed in the left column of Table~\ref{tab:TZ_RVfits}.}
         \label{Fig_TZ_Single_RV}
   \end{figure}

As for \tyc, we started by fitting a single-planet model to the RV data of TZ\,Ari. Here we used the combined data set from HIRES, HARPS, and CARMENES (see Sect.\ref{Spec}), and allowed for individual RV zero points and jitter terms for the three instruments. The best-fit parameters are listed in the left column of Table~\ref{tab:TZ_RVfits}, and the data together with the best fit are shown in Fig.~\ref{Fig_TZ_Single_RV}. A highly eccentric Keplerian with period $\sim 770$\,d provides a reasonable fit to all data, albeit with a rather large seemingly white-noise scatter. As we will show in the next section, however, this scatter does bear the imprint of the 1.96-day rotation period, and can readily be modeled and explained by it. We thus ascribe the 770\,d signal to a planet, TZ\,Ari\,b.

\subsubsection{Gaussian process model}
\label{TZ_GP_model}

\begin{figure}
   \centering
   \includegraphics[width=\hsize]{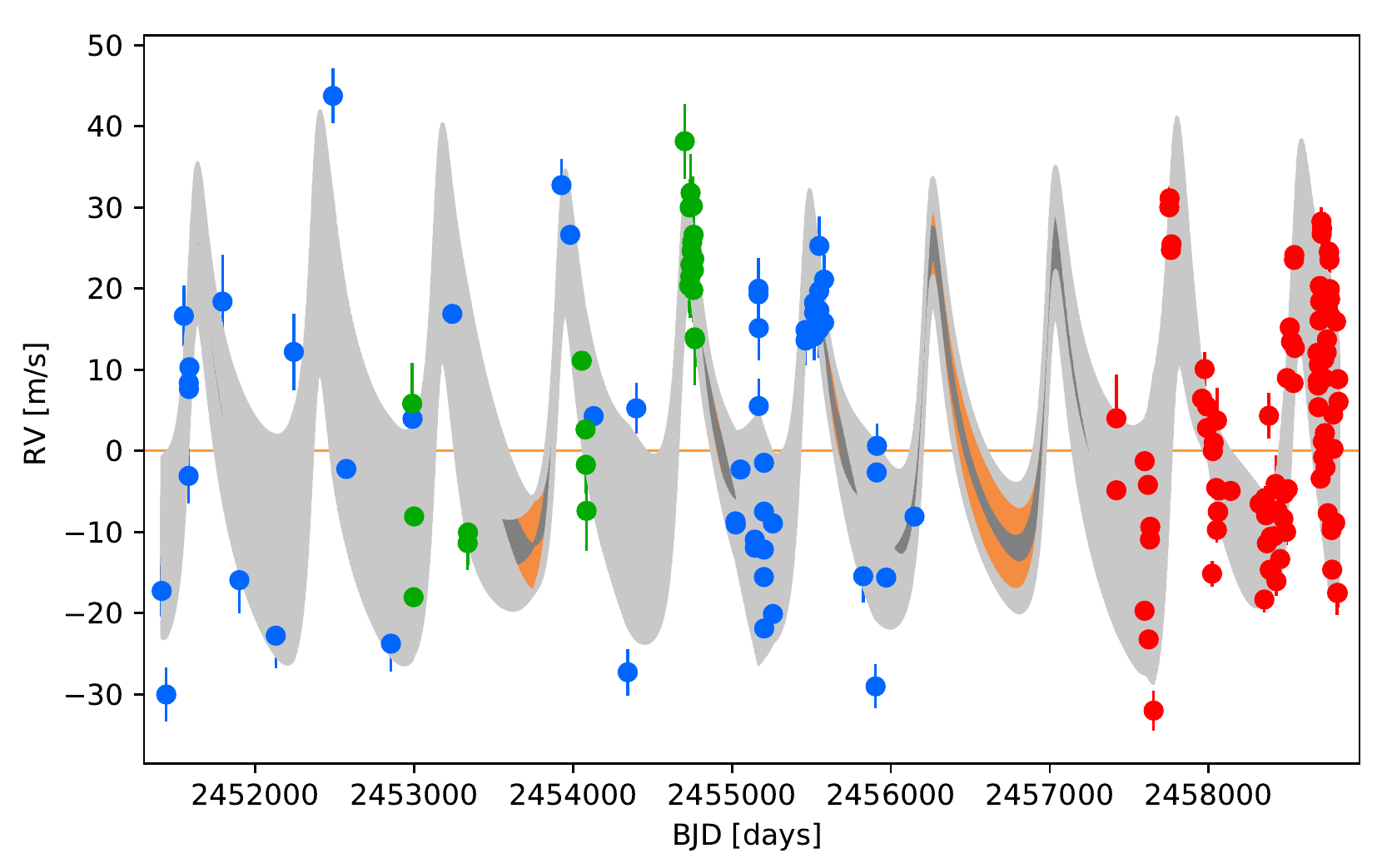}
      \caption{Gaussian process model fit to the RV data of TZ\,Ari from HIRES (blue), HARPS (green), and CARMENES VIS (red). The orbital parameters are listed in the right column of Table~\ref{tab:TZ_RVfits}. The rotational modulation of the RV is far too fast to be seen at any reasonable screen or print resolution; the full range of this variation is therefore seen as a grey band. We provide a zoom into a short section in Fig.~\ref{Fig_TZ_GP_RV_Zoom}.}
         \label{Fig_TZ_GP_RV}
   \end{figure}

\begin{figure}
   \centering
   \includegraphics[width=\hsize]{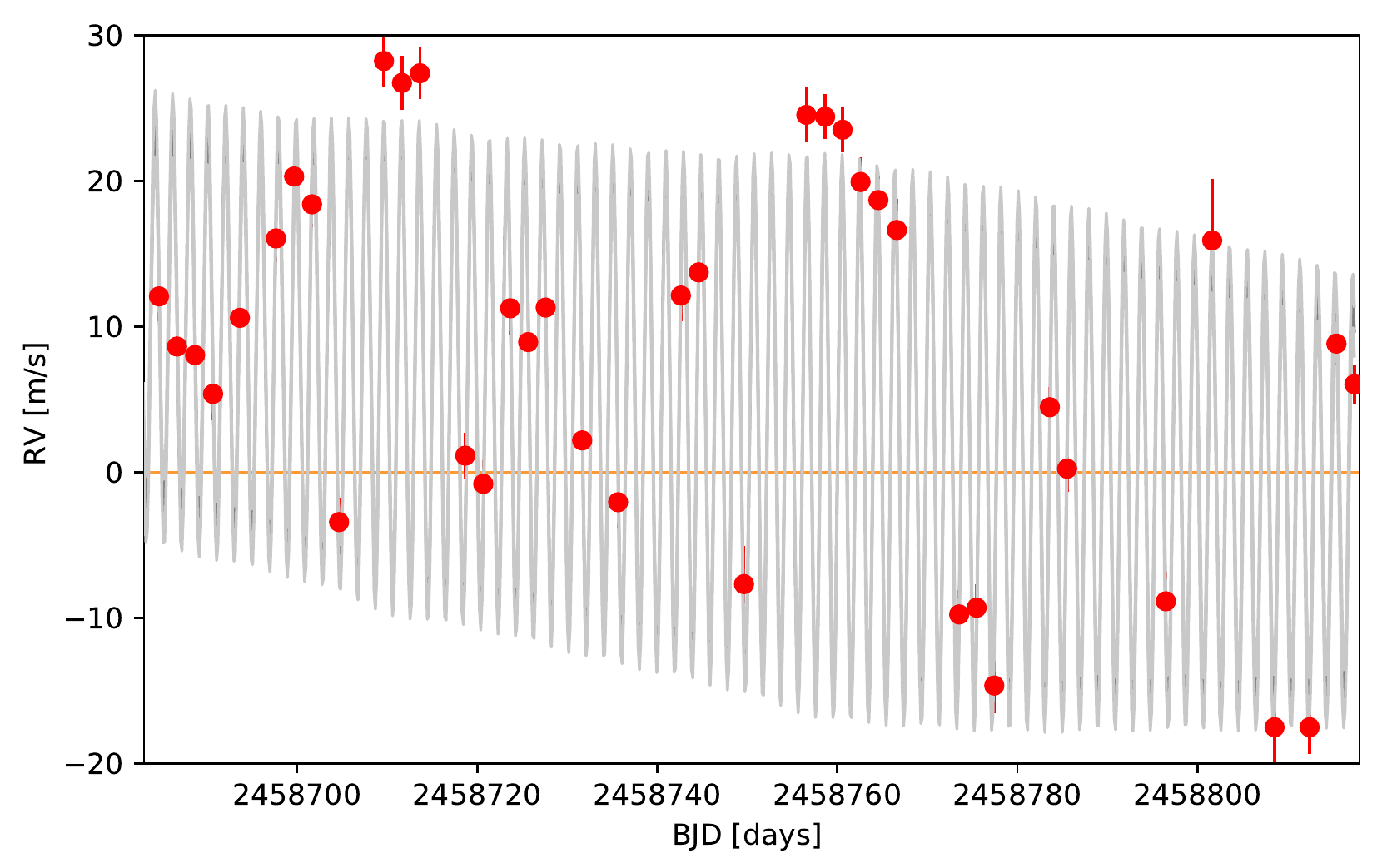}
      \caption{Zoom onto the Gaussian process model fit to the RV data of TZ\,Ari shown in Fig.~\ref{Fig_TZ_GP_RV}. The fast modulation with a period of 1.96\,d is due to the stellar rotation. The long-term downward trend is part of the 770-day planetary orbit.}
         \label{Fig_TZ_GP_RV_Zoom}
   \end{figure}

\begin{figure}
   \centering
   \includegraphics[width=\hsize]{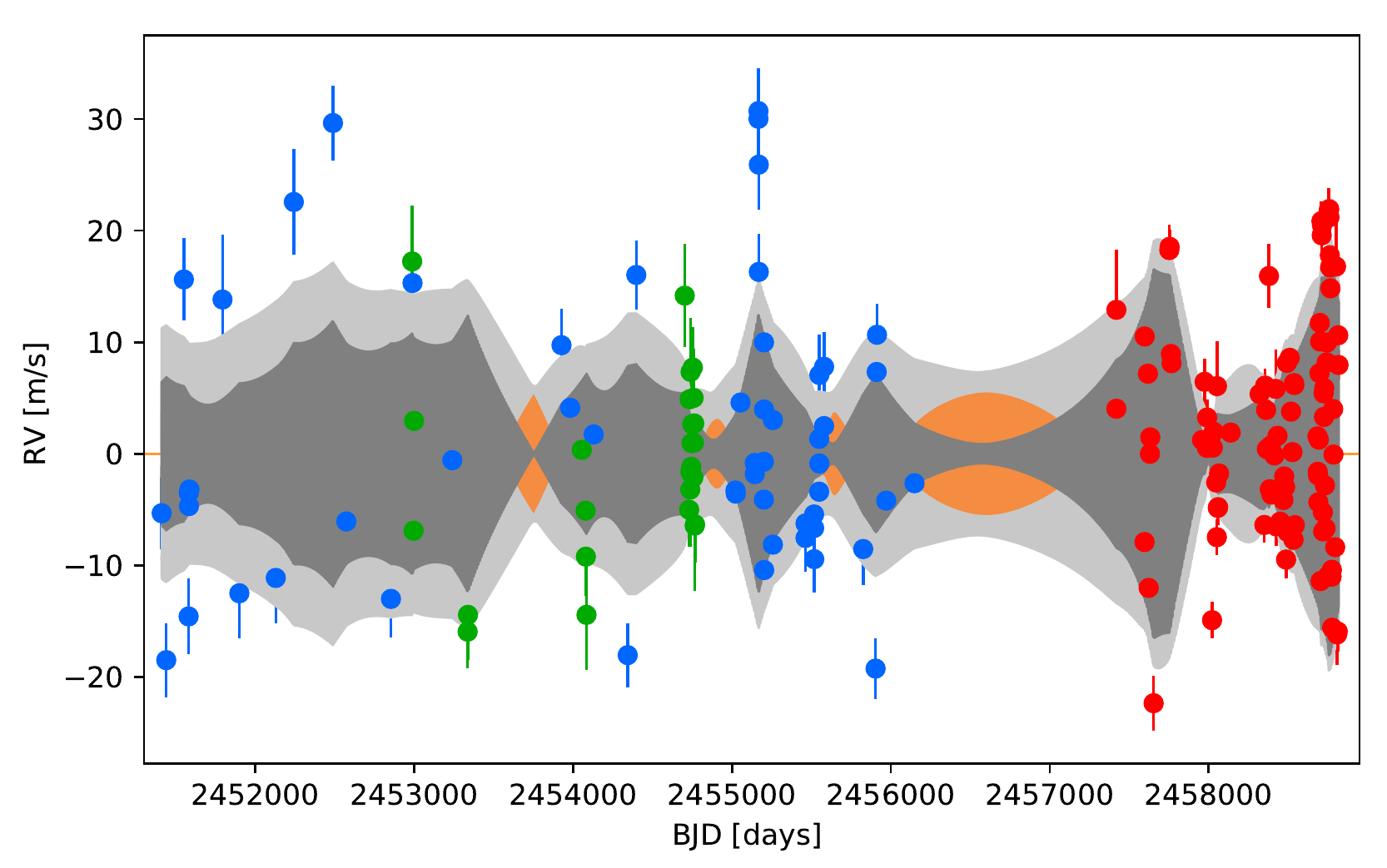}
      \caption{Gaussian process model fit to the RV data of TZ\,Ari as in Fig.~\ref{Fig_TZ_GP_RV}, but with the planet orbit fit subtracted. Together with Fig.~\ref{Fig_TZ_GP_RV_Zoom}, this figure shows that the SHO model reproduces the $1.96$-day quasi-periodic variations very well, without introducing any power on longer time scales, which could affect the best-fit planet parameters.}
         \label{Fig_TZ_GP_RV_Res}
   \end{figure}

\begin{figure}
   \centering
   \includegraphics[width=\hsize]{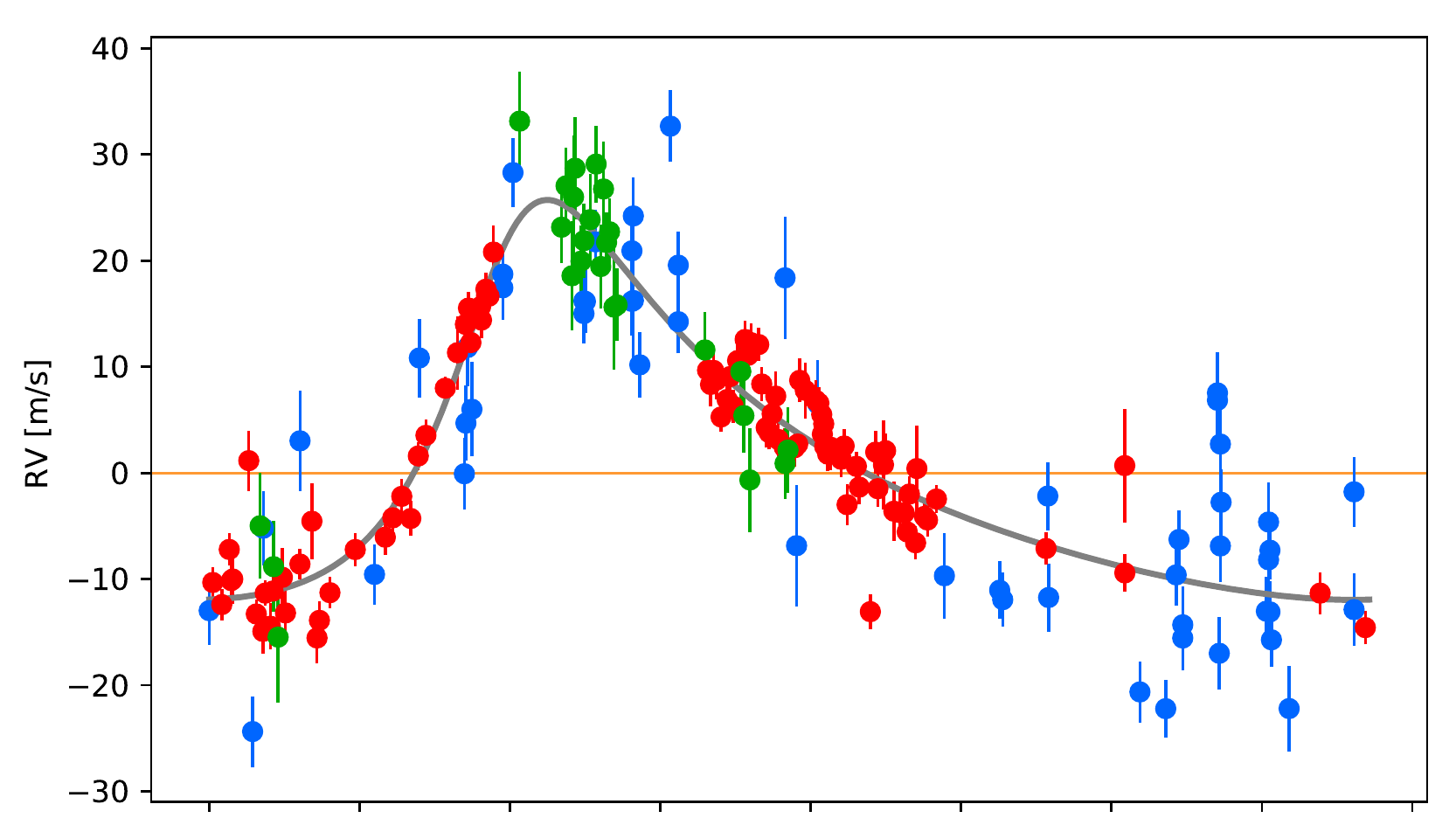}
   \includegraphics[width=\hsize]{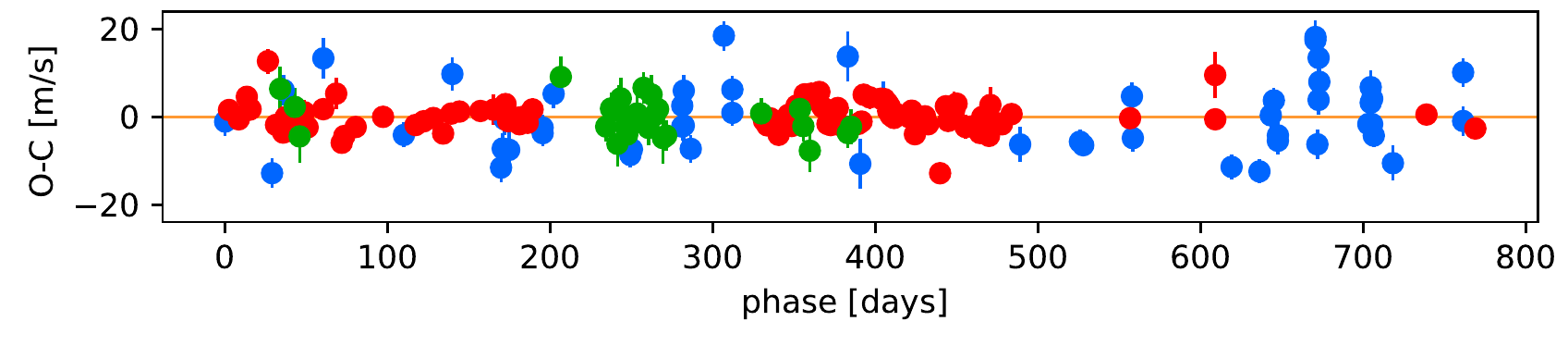}
      \caption{Gaussian process model fit to the RV data of TZ\,Ari as in Fig.~\ref{Fig_TZ_GP_RV}, but with the Gaussian process model of the stellar variability subtracted, and folded to the period of TZ\,Ari\,b (771.36\,d).}
         \label{Fig_TZ_GP_RV_Folded}
   \end{figure}

\begin{figure}
   \centering
   \includegraphics[width=\hsize]{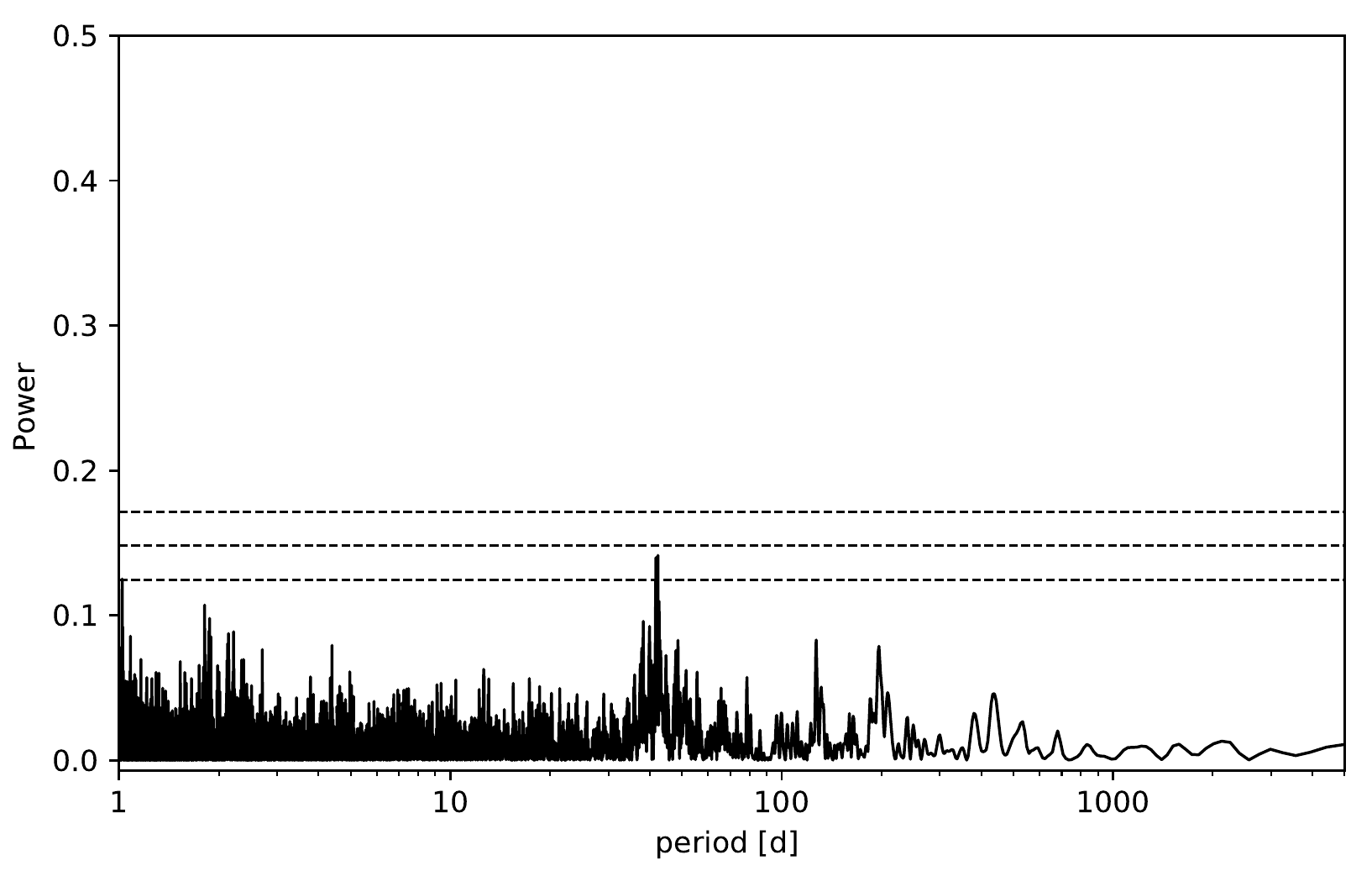}
      \caption{GLS periodogram of the residuals of the GP model for TZ\,Ari. False-alarm probabilities of 0.1, 0.01, and 0.001 are shown as dashed horizontal lines. The scales of the axes are identical to Fig.~\ref{Fig_TZ_Periodogram} for better comparison.}
         \label{Fig_TZ_Periodogram_Res}
   \end{figure}

Proceeding as with \tyc{}, we also modeled the RV data of TZ\,Ari with a GP, using the priors in Appendix~\ref{GP_priors}. Here we let 700 walkers run for 1,000 steps each, discarded the first 500, and retained 118,702 samples for the computation of the model parameters and their confidence ranges. The results are shown in Figs.~\ref{Fig_TZ_GP_RV} to \ref{Fig_TZ_GP_RV_Folded} and listed in the right column of Table~\ref{tab:TZ_RVfits}, and more detailed information is again shown in Appendix~\ref{cornerplots}. Comparing the two columns in Table~\ref{tab:TZ_RVfits}, there is good overall agreement between the two models, but with a $\sim 10$\,\% lower value of $K$ for the GP model. A closer inspection of the data reveals a correlation between $K$ and the zero point of the HARPS RVs, which are clustered around the maximum of the RV curve (see Fig.~\ref{Fig_TZ_GP_RV_Folded}, also noting the elongation of the distribution in the field ``$K_b$, RV off$_3$'' in Fig.~ \ref{Fig_TZ_GP_corner}). As the GP yields a fully consistent model (apparent in Fig.~\ref{Fig_TZ_GP_RV_Folded}) and as this model is again strongly favored over the single-planet model ($\Delta{\rm BIC} = -152$), we adopted the corresponding parameters for TZ\,Ari\,b as listed in the right-hand column of Table~\ref{tab:TZ_RVfits}.

The residuals of the GP model for TZ\,Ari are shown in Fig.~\ref{Fig_TZ_Periodogram_Res}. The only peak exceeding the 10\,\% false-alarm probability threshold, at 42.3\,d, is probably an alias of the second harmonic of the rotation ($1/42.3 + 1 \approx 2/1.96$). As in the previous case, we conclude that all available data on TZ\,Ari can be modeled with a single planet plus rotational modulation.

\subsubsection{Considering whether the 770-day signal could be due to activity cycles}
\label{act_cycle}

While the modeling of the 770-day RV signal presented in the previous sections is fully compatible with a planetary origin, we also conducted additional checks to address the possibility that it could instead be caused by stellar activity cycles:
\begin{itemize}
  \item An RV signal induced by stellar activity would likely have a higher amplitude at shorter wavelengths, because it would be related to temperature variations. We therefore compared the amplitudes of the signal from the visible and infrared spectrographs of CARMENES with each other, but found agreement between the respective amplitudes (21.5\,m\,s$^{-1}$ and 21.0\,m\,s$^{-1}$, respectively).
  \item The combined RV data set (see Fig.~\ref{Fig_TZ_GP_RV}) covers ten cycles without any noticeable change of either amplitude or phase. This is also apparent from the near-perfect folding of the RVs with phase (see Fig.~\ref{Fig_TZ_GP_RV_Folded}). In contrast, activity cycles might vary in strength or duration, as observed in the Sun.
  \item It might be expected that activity cycles would manifest themselves through, for example, the variability of the H$\alpha$ line. However, the periodograms of CRX, dLW, and spectroscopic indicators described in Sect.~\ref{Rot} and shown in Appendix~\ref{Indicators} do not show any significant power near 770\,d.
  \item The effect of activity cycles on the RVs would likely not only be a smooth periodic variation, but also a modulation of the short-term scatter with the phase of the cycle. We therefore subtracted the single-Keplerian fit of Fig.~\ref{Fig_TZ_Single_RV} -- meant here to represent the smooth RV variations -- from the data, and computed a GLS periodogram of the absolute values of the residuals. No excess power near 770\,d was found, indicating that the RV scatter does not depend on the phase of the long-term variations.
  \item No long-term photometric variations are observed in our SuperWASP photometry or in publicly available data from MEarth \citep{Berta2012}, with an upper limit of $\sim 5$\,mmag on periods near 770\,d. This is not a very constraining limit compared to photometric variations of the Sun, which are only on an order of 1\,mmag, but achieving this precision with ground-based photometric monitoring would be very difficult.
\end{itemize}
Although none of these arguments by itself excludes activity cycles as a possible cause of the 770-day signal, the preponderance of the evidence strongly points to a planetary origin.

\subsubsection{Considering a second planet in the system}
\label{Second_planet}

\begin{figure}
   \centering
   \includegraphics[width=\hsize]{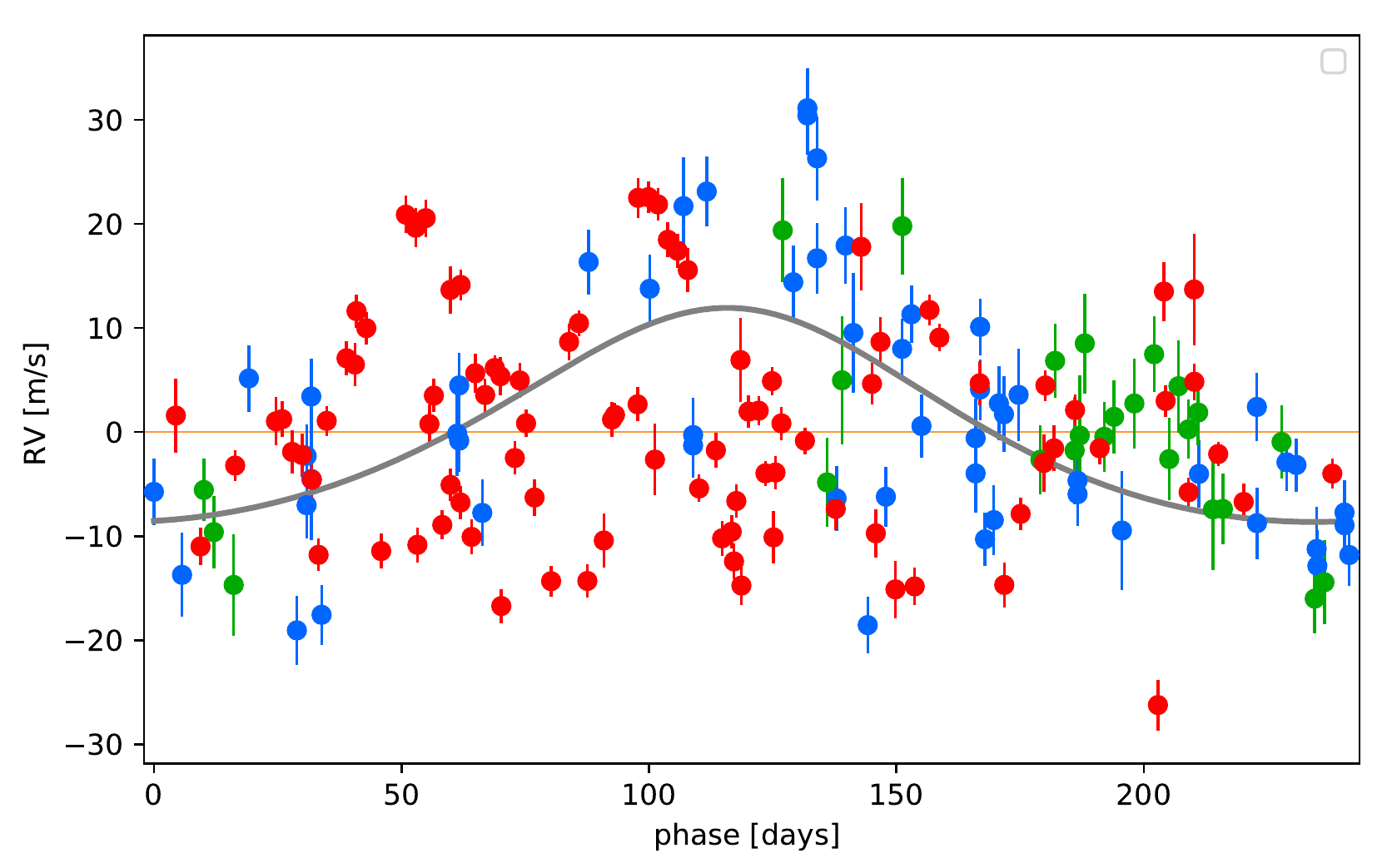}
      \caption{Two-planet fit for TZ\,Ari from \citet{Feng2020}. Their model of the 770\,d-planet was subtracted, and the residuals phase-folded to the period of the putative second planet, 241.59\,d. While the HIRES (blue) and HARPS (green) data weakly support this planet candidate, the CARMENES VIS (red) data rule out its existence.}
         \label{Feng_Planet2}
   \end{figure}

The 770\,d RV periodicity was already detected in the HIRES RV time series and was classified as a ``signal requiring confirmation'' by \citet{Butler2017}. In the combined HIRES and HARPS data sets, \citet{Tuomi2019} identified three planet candidates with orbital periods of 773.4\,d, 1.932\,d, and 243.1\,d; the first and third of these are also listed by \citet{Feng2020}. The first of these candidates has now been established as a planet thanks to the larger, more precise, and better sampled CARMENES dataset, and to the arguments presented in Sect.~\ref{act_cycle}. The second candidate is definitely spurious. It corresponds to the 1.96\,d signal, which can firmly be attributed to rotational modulation because it is chromatic, involves variations in the line shape (dLW), and is accompanied by photometric variability, as well as periodic changes of the line indices (see Sect.~\ref{Rot_TZ}).

To assess the reality of the 240\,d planet candidate, we first note that the two-planet model of \citet{Feng2020} is incompatible with the CARMENES data (see Fig.~\ref{Feng_Planet2}). For a more thorough analysis, we extend the model of Sect.~\ref{TZ_GP_model} by adding a second planet with a period constrained to $P \in [240,245]$\,d, with all other parameters left free. The best-fit amplitude is found to be $K = 0.95 \pm 1.05$\,m\,s$^{-1}$, and thus compatible with zero. This demonstrates that there is no evidence for a planet in this period range in the combined data set.

We conclude that on the basis of the CARMENES data, we can firmly establish TZ\,Ari\,b as a bona fide planet and we demonstrate that the previously suspected additional planet candidates are spurious.

\section{Discussion}
\label{Discussion}

\subsection{\tyc\,b}
\label{TYC_b}

The planet \tyc\,b is about 10\,\% more massive than Saturn; the eccentricity of its orbit is compatible with zero. With an insolation between that of Jupiter and Saturn (more precisely, about 2.5 times that of Saturn), \tyc\,b could be physically quite similar to the gas giants in the Solar System.
The system does in fact resemble a number of others with an early M host and a single Saturn-like planet in a long-period orbit (see Sect.~\ref{CARMENESGiants} and Fig.~\ref{Fig_Mass_Period}).

\subsection{TZ\,Ari\,b}
\label{TZ_b}

The companion of TZ\,Ari is much more uncommon: it is only the second known giant planet orbiting a host with $M_\star \leq 0.3\,M_\odot$, along with GJ\,3512\,b \citep{Morales2019} and not counting the candidate GJ\,3512\,c. The unusual nature of the GJ\,3512 and TZ\,Ari systems is quite apparent in a diagram of planet mass, $m,$ versus host mass, $M_\star$, where they are separated from other known systems by a gap of a factor of $\sim 2$ in $M_\star$ or a factor of $\sim 5$ in $m$ \citep[see Fig.~1 in][and Fig.~\ref{Fig_Mass_Period}]{Quirrenbach2020}. The orbits of TZ\,Ari\,b and GJ\,3512\,b are rather similar: While the former has a longer period than the latter (771\,d vs.\ 204\,d), both have rather large and nearly identical eccentricity (0.46 and 0.44, respectively).

As noted in Sect.~\ref{TZprop}, the observational limit on $v \sin i_\star$ leads to $P_{\rm rot} / \sin {i_\star} > 4.15$\,d. Combining this with the rotation period of 1.96\,d, we conclude that the stellar rotation axis must be rather inclined, $\sin{i_\star} < 0.47$. This means that TZ\,Ari\,b must have either a rather misaligned orbit or a mass of at least 0.45\,$M_{\rm Jup}$ (if aligned), which is twice as great as the minimum mass, $m \sin i$. This question will certainly be resolved once the full $Gaia$ data, including astrometric orbit solutions, are published.

TZ\,Ari\,b is also among the closest giant exoplanets to the Earth, which means that its angular orbital semi-major axis of $0 \farcs 27$ is rather large. With a radius that should be about ten times larger than the Earth's, the contrast between TZ\,Ari\,b and its host is two orders of magnitudes more favorable than that of a terrestrial planet, which makes it a very suitable target for characterization by a future direct imaging mission. This will be particularly interesting because of the eccentric orbit, which gives rise to periodic variations of the insolation by a factor of more than seven.

The orbit-averaged insolation of TZ\,Ari\,b is only about 30\,\% that of Saturn. Its energy balance will therefore remain dominated by internal heat beyond an age of 10\,Gyr \citep{Linder2019}. Consequently, the effective temperature of TZ\,Ari\,b will slowly approach its equilibrium value $T_{\rm eff} \approx 65$\,K, with a present value likely below 100\,K. It is therefore no surprise that the planet was not discovered in the 8.7\,$\mu$m survey of \cite{Gauza2021}. Searches for reflected light will be more promising, with an expected contrast of order $3 \cdot 10^{-8}$ between the planet and its host.

\subsection{Giant planets in the CARMENES M-dwarf sample}
\label{CARMENESGiants}

\begin{table*}
\centering
\small
\caption{Giant planets ($m \sin i \geq 0.5\,M_{\rm Sat} \approx 0.15\,M_{\rm Jup}$) in the CARMENES sample.} \label{tab:giant_planets}
\begin{tabular}{lcccccr}
\hline\hline
\noalign{\smallskip}
Name & $M_\star$ $[M_\odot]$ & $m \sin i$ $[M_{\rm Jup}]$ & $m \sin i/M_\star$ & $P$ [d] & $e$ & Reference\\
\hline
\noalign{\smallskip}
GJ\,3512\,b & 0.123 & 0.463 & $3.59\times 10^{-3}$ & 203.59 & 0.44 & M19\\
TZ\,Ari\,b & 0.150 & 0.213 & $1.36\times 10^{-3}$ & 771.36 & 0.46 & This work\\
GJ\,876\,c & 0.327 & 0.726 & $2.12 \times 10^{-3}$ & 30.126 & 0.25 & T18\\
GJ\,876\,b & 0.327 & 2.288 & $6.68 \times 10^{-3}$ & 61.082 & 0.03 & T18\\
GJ\,1148\,b & 0.354 & 0.304 & $0.82 \times 10^{-3}$ & 41.380 & 0.38 & T20a\\
GJ\,1148\,c & 0.354 & 0.227 & $0.61 \times 10^{-3}$ & 532.64 & 0.38 & T20a\\
GJ\,179\,b & 0.357 & 0.82 & $0.36 \times 10^{-3}$ & 2288 & 0.21 & H10\\
GJ\,15\,Ac & 0.391 & 0.152 & $0.37 \times 10^{-3}$ & 7025 & -- & T18\\
GJ\,849\,b & 0.468 & 0.884 & $1.80 \times 10^{-3}$ & 1924 & 0.04 & F15\\
GJ\,849\,c & 0.468 & 0.916 & $1.87 \times 10^{-3}$ & 5520 & 0.09 & F15\\
\tyc{}\,b & 0.498 & 0.328 & $0.63\times 10^{-3}$ & 691.90 & 0.05 & This work\\
GJ\,649\,b & 0.514 & 0.317 & $0.59\times 10^{-3}$ & 598.3 & 0.30 & J10b\\
\noalign{\smallskip}
\hline
\end{tabular}
\tablebib{
    M19 \cite{Morales2019}; T18: \cite{Trifonov2018}; T20a: \cite{Trifonov2020a}; H10: \cite{Howard2010}; F15: \cite{Feng2015}; J10: \cite{Johnson2010b}. Stellar masses have been updated from \cite{Schweitzer2019}, and the derived planet masses corrected according to $m \sin i \propto M_\star^{2/3}$.
}
\end{table*}

The CARMENES survey was designed to provide statistical information on planetary systems of low-mass stars. In fact, it covers the mass range from the hydrogen burning limit at 0.07\,$M_\odot$ to the upper end of the M spectral class near 0.6\,$M_\odot$ rather uniformly \citep[see Fig.~2 in][]{Reiners2018}. While a thorough statistical analysis has to await the completion of the survey, it is already possible to draw preliminary conclusions on the occurrence rate of giant planets in the sample. For the following discussion, we adopt a somewhat arbitrary definition of $m \sin i \geq 0.5\,M_{\rm Sat} \approx 0.15\,M_{\rm Jup}$ for this class. The twelve planets in the CARMENES sample of 387 M dwarfs (including those found independently) fulfilling this criterion are listed in Table~\ref{tab:giant_planets}. The entries in the table immediately provide a lower limit on the number of the giant planets in this sample. On the other hand, we can roughly estimate the completeness of this list as follows: So far, 160 stars have been observed at least 30 times during the survey, with RVs taken over a time span of more than 1\,yr. A giant planet with $m \sin i \geq 0.5 M_{\rm Sat}$ in a circular orbit with $P \leq 2$\,yr around a star with $M_\star \leq 0.5\,M_\odot$ gives rise to a signal of $K \geq 5.4$\,m\,s$^{-1}$ and would thus have easily been found in any of these 160 RV curves, unless its presence is masked by strong stellar activity, which may be the case for up to $\sim 25\,\%$ of the stars \citep{Tal-Or2018}. We thus conclude that the completeness of Table~\ref{tab:giant_planets} for periods up to two years is at least $0.75 \cdot (160/387) \approx 33\,\%$, and probably substantially higher because many additional stars have been observed already more than 20 times. The completeness drops towards longer periods because $K \propto P^{-1/3}$ and because many stars have now been monitored for two to three years.

From an inspection of Table~\ref{tab:giant_planets} we can draw a few direct conclusions about the population of giant planets orbiting M dwarfs:
\begin{itemize}
  \item Giant planets exist around stars of all masses down to at least 0.12\,$M_\odot$, with orbital periods ranging from a few weeks to at least two decades.
  \item While planets with $m \sin i \geq 1\,M_{\rm Jup}$ appear to be rare (with GJ\,876\,b being the only one in the sample), ``mass ratios'' $m \sin i / M_\star$ larger than that of Jupiter and the Sun are actually quite common.
  \item The observed orbital eccentricities range from values compatible with zero to $e = 0.46$.
  \item At least three of the nine known systems in the CARMENES sample have multiple giant planets.
  \item The inferred occurrence rate of M dwarfs hosting giant planets with periods up to 2 years is 2\,\% to 6\,\%, with the limits representing the assumptions that the table is 100\,\% or only 33\,\% complete. This is somewhat lower than the corresponding number for FGK stars (roughly 10\,\%, see e.g.\ \citealt{Fernandes2019, Wittenmyer2020} and Table~1 in \citealt{Winn2015}), but the estimates come closer together if $m \sin i / M_\star$ is considered instead of $m \sin i$.
\end{itemize}
As a cautionary remark, we note that the mass range below $M_\star \sim 0.3\,M_\odot$ remains poorly explored, with good RV coverage for only a few dozen stars resulting in two detections, namely GJ\,3512\,b and TZ\,Ari\,b. Here, the completion of the CARMENES survey, and its extension that began in 2021, will provide statistical information on more than 350 nearby M stars. We are aiming at obtaining uniform detection limits, and we will perform detailed injection-retrieval experiments, extending the analysis of 71 CARMENES stars by \citet{Sabotta2021} to the whole sample.

\subsection{Giant planets orbiting low-mass stars}
\label{LMSGiants}

\begin{figure}
   \centering
   \includegraphics[width=\hsize]{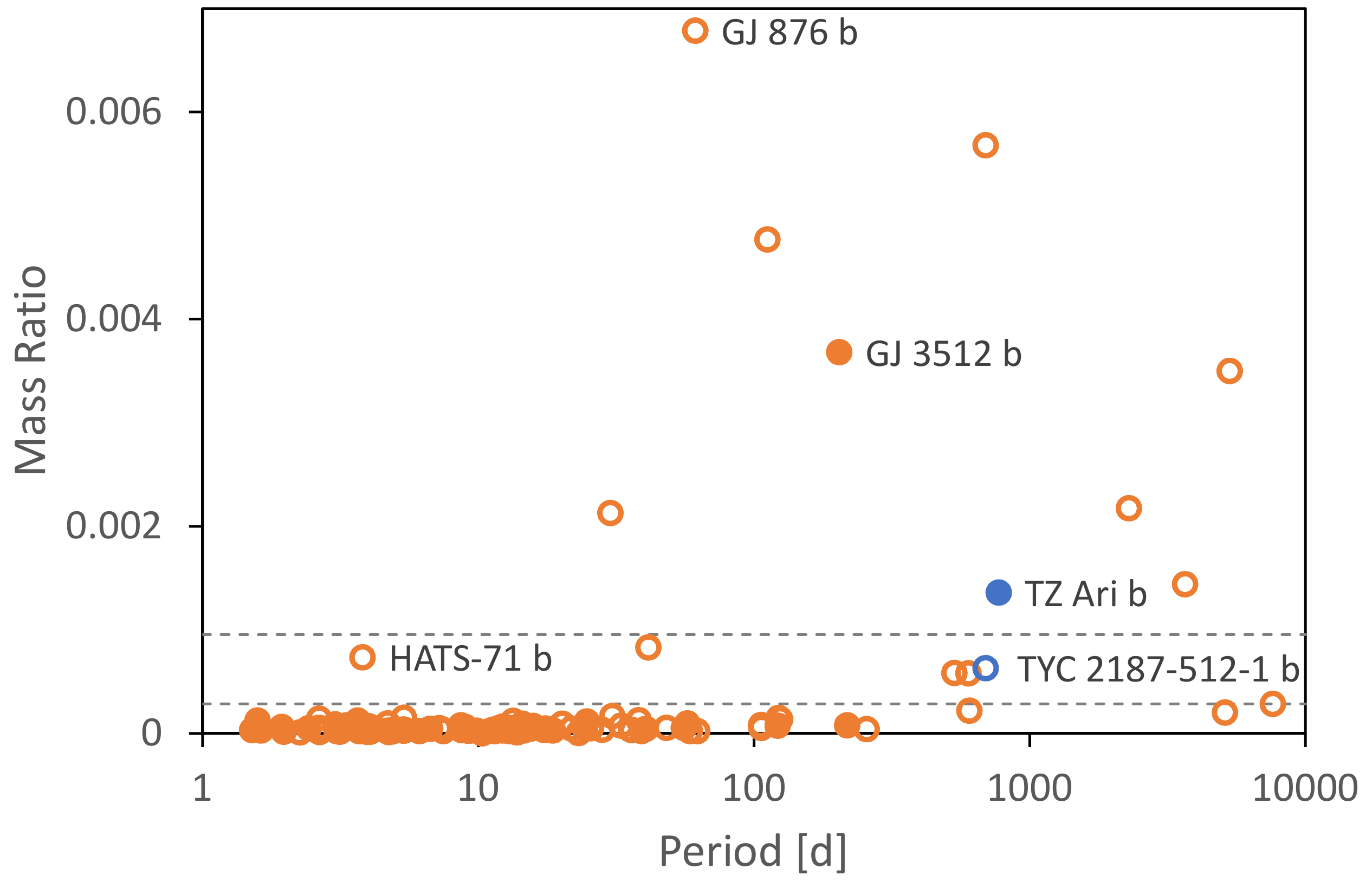}
      \caption{Mass ratio-period relation for planets orbiting low-mass stars. The y-axis shows the mass ratio $m/M_\star$ (for transiting planets) or $m \sin i / M_\star$ (for non-transiting planets). The dashed horizontal lines indicate the mass ratios of Saturn and Jupiter compared to the Sun. The open symbols designate planets orbiting hosts with $0.3\,M_\odot < M_\star \leq 0.55\,M_\odot$, the filled symbols are for host masses $M_\star \leq 0.3\,M_\odot$. Objects with small mass ratios, that is,\ terrestrial planets and super-Earths, are included for illustration, but not discussed further in the present paper. The data were compiled from the NASA Exoplanet Archive.}
         \label{Fig_Mass_Period}
   \end{figure}

To get a broader overview of the giant planet population orbiting low-mass stars, we compiled a list of these objects from the NASA Exoplanet Archive\footnote{\url{https://exoplanetarchive.ipac.caltech.edu}}. We note that objects with masses ranging from $\sim 1\,M_{\rm Jup}$ to the low-mass stellar regime have nearly identical diameters, and therefore objects discovered by transit surveys without mass measurements are of little value for the following discussion. Starting from a database query for all planets with masses (or $m \sin i$) known from radial velocities or transit timing variations, and with host mass $M_\star \leq 0.55\,M_\odot$, we disregarded a few host stars with $T_{\rm eff} > 4300$\,K (and thus obviously dubious mass estimates). We computed a ``mass ratio'' for each planet, which is either $m/M_\star$ or $m \sin i / M_\star$, depending on whether $\sin i$ is known or not. As the expectation value for a random orientation of the orbits $\langle \sin i \rangle = \frac{\pi}{4} = 0.79$, which is rather close to one, we make no further attempt to place limits on $\sin i$ for the non-transiting planets, and use $m \sin i$ and the mass ratio as defined here as good proxies for the underlying masses and ``true'' mass ratios.

Figure~\ref{Fig_Mass_Period} shows the mass ratio for this sample of 112 planets, together with \tyc\,b and TZ\,Ari\,b, as a function of their orbital period.
This figure makes it apparent that: ($i$) no ``hot Jupiter'' with a period of $P \le 10$\,d and host mass of $M_\star \leq 0.55\,M_\odot$ has been found thus far and that ($ii$) giant planets with longer periods of $P \ge 30$\,d and low-mass hosts are not uncommon. The two discoveries from the CARMENES Survey GJ\,3512\,b and TZ\,Ari\,b demonstrate that this second statement remains true down to host masses of $M_\star \approx 0.15\,M_\odot$, as mentioned in Sect.~\ref{CARMENESGiants}.

The period distribution of giant planets discovered in RV surveys of FGK stars shows a lack of objects in the range between 10 and 100\,d, sometimes dubbed the ``period valley'' \citep{Udry2003, Santerne2016}, with higher occurences at shorter and at longer periods. No such valley is apparent in Fig.~\ref{Fig_Mass_Period}.
The occurrence rate of hot Jupiters in samples of AFGK stars observed by {\it Kepler} and {\it TESS} is about 0.4\,\%, with a slight, statistically insignificant tendency towards higher values at lower stellar masses \citep{Zhou2019}. This possible trend does not seem to continue into the M dwarf range, but the current absence of hot Jupiters from the M dwarf samples could also just be due to small number statistics. In Fig.~\ref{Fig_Mass_Period}, we also mark HATS-71\,b, for which a mass of $0.37 \pm 0.24\,M_{\rm Jup}$ and a radius of $1.024 \pm 0.018\,R_{\rm Jup}$ have been measured \citep{Bakos2020}. These values are indicative of a hot and somewhat inflated Saturn, but also compatible with a hot Jupiter. More mass measurements of gas giants transiting hosts with masses $M_\star \leq 0.5\,M_\odot$, together with RV surveys of larger M dwarf samples, are needed to ascertain whether the apparent differences between low-mass stars and FGK stars are genuine or due, rather, to the present small-number statistics.

\subsection{Astrometric constraints}
\label{Astrometry}

For stars with known parallax $\varpi$, the spectroscopic orbital parameters can directly be converted into a lower limit for the astrometric signature $\alpha$, namely, the semi-major axis of the orbital ellipse projected onto the sky \citep{Reffert2011}\footnote{Note that there is a typo in Eqn.~2 of \citet{Reffert2011}, where a square root was lost in the conversion from Eqn.~1.}:
\begin{equation}\label{astrom_sign}
  \alpha_{\rm min} = \frac{K \cdot P \cdot (1 - e^2) \cdot \varpi}{1\,{\rm au}\cdot 2 \pi} ~~.
\end{equation}
For \tyc\,b and TZ\,Ari\,b, we get $\alpha_{\rm min} = 49.1$\,$\mu$as and $\alpha_{\rm min} = 235$\,$\mu$as, respectively, well within the capabilities of the $Gaia$ mission. It is indeed expected that $Gaia$ will provide a rather complete census of giant planets orbiting dwarf stars in the Solar neighborhood with intermediate periods \citep{Sozzetti2014,Reyle2021}. This will likely include all planets in Table~\ref{tab:giant_planets}, except GJ\,1148\,b and GJ\,15\,Ac, which have a very small astrometric signature and a very long period, respectively. Since astrometric orbit solutions include the inclination, the dynamical masses of these planets will be known as soon as the $Gaia$ orbits are published.

Even astrometric non-detections can provide useful constraints on spectroscopic orbits, as it is always possible to infer a lower limit to $\sin i$ from them. This requires access to the individual astrometric measurements, which are not yet available for $Gaia$. However, the published data releases contain information on the astrometric orbit in the form of the parameter $\epsilon_i$, the ``excess source noise'' quantifying the residuals from the standard five-parameter astrometric fit \citep{Lindegren2012,Lindegren2016}. While caution has to be exercised in interpreting $\epsilon_i$ for individual stars, the presence of a companion with a significant signature will usually lead to a correspondingly high value of $\epsilon_i$. An inspection of $\epsilon_i$ in $Gaia$ EDR3 for the stars in Table~\ref{tab:giant_planets} reveals that they are all significantly different from zero, but never larger than 1\,mas. This means that it is highly unlikely that any face-on stellar binaries have crept into this list. In addition, even inclinations making the companions brown dwarfs are highly unlikely ($\sin i < 0.07$, corresponding to $p < 0.25\,\%$ in all but one cases), so that the tabulated objects can safely be considered gas giant planets.

\subsection{Planet formation scenarios}
\label{Formation}

The formation of rocky planets orbiting very low-mass stars such as TRAPPIST-1 \citep{Gillon2017} and Teegarden's Star \citep{Zechmeister2019} can be explained with models based on the planetesimal accretion \citep{Miguel2020} or pebble accretion \citep{Schoonenberg2019} paradigms. It is very difficult, however, to incorporate the formation of gas giants in these scenarios \citep{Morales2019}, because a high migration rate is expected for rocky cores forming in the disks of late M dwarfs. Artificially reducing the speed of type~I migration may point towards a solution of this conundrum \citep{Burn2021}, but it is also plausible that the dominant formation channel of these planets is through gravitational instabilities in their natal disk \citep{Kratter2016}. This possibility was explored by \citet{Mercer2020}, who found that disk fragmentation is a viable pathway provided that the initial disk is sufficiently large, namely, about 30\,\% of the stellar mass. In this context, it is worth noting that the ``mass ratio,'' $m \sin i/M_\star$, of objects such as GJ\,3512\,b and TZ\,Ari\,b is comparable to that of a low-mass brown dwarf around a Sun-like star, or to the pair of brown dwarfs orbiting the K giant $\nu$\,Oph \citep{Quirrenbach2019}, which may also have formed via disk instabilities. Around solar-type stars, there are indeed indications for two separate giant planet populations whose occurrence rates have different dependencies on stellar metallicity: planets below $\sim 4\,M_{\rm Jup}$ follow a positive metallicity-occurrence relation, whereas planets above this mass are found around host stars with a wide range of metallicities \citep{Santos2017,Schlaufman2018}. The former trend is expected from a core accretion scenario, whereas the latter is consistent with disk fragmentation.
More extensive surveys providing better statistics of the giant planet population over a large range of host star masses would help clarify this issue. Trends of planet occurrence rate with host star metallicity could also help to distinguish between different formation pathways. This possibility lends added urgency to a resolution of the remaining uncertainties in the determination of [Fe/H] in M dwarfs (see the discussions in \citealt{Passegger2022} and \citealt{Marfil2021}). An alternative route would be the exploitation of Vanadium as a proxy of M dwarf metallicities \citep{Shan2021}.

The high eccentricity of the orbits of GJ\,3512\,b and TZ\,Ari\,b could be a result of past planet-planet interaction in a multiple system \citep{Ida2013, Carrera2019}. The RV data of GJ\,3512 do indeed show a long-term modulation indicative of a second outer planet candidate with a period of at least four years \citep{Morales2019}. In contrast, there is no indication of any long-period trend in the RVs of TZ\,Ari (see Figs.~\ref{Fig_TZ_GP_RV} and \ref{Fig_TZ_GP_RV_Res}). The time series covers 20 years, but the intrinsic RV variability of the star limits the ability to detect long-period planets. Because of the proximity of the star, a search for an acceleration in the $Gaia$ astrometric data set will be the most sensitive test for the presence of a third body in the system.

\section{Conclusions}
\label{Conclusions}

We present two Saturn-mass planets in wide orbits around M dwarfs, \tyc\,b and TZ\,Ari\,b, whose parameters can be established from RV measurements even in the presence of a rather strong signature of stellar rotation in these data. While \tyc\,b, which orbits an early M star, is not unusual as a number of other planets with similar properties are known, TZ\,Ari\,b is only the second confirmed giant planet orbiting a star with mass $M_\star \leq 0.3\,M_\odot$. In fact, the mass of TZ\,Ari is only $0.15\,M_\odot$, similar to that of GJ\,3512, the other low-mass host of a gas giant planet. These two systems occupy an extreme region of the planet mass / host mass parameter space, where disk fragmentation appears to be a more likely formation mechanism than build-up of a core through planetesimal or pebble accretion.

The orbits of both GJ\,3512\,b and TZ\,Ari\,b are rather eccentric ($e = 0.44$ and $e = 0.46$, respectively), which could be explained if the planets reside in multiple systems. Whereas a candidate outer planet, GJ\,3512\,c, has been identified in the first case, there is no indication for an additional companion of TZ\,Ari in the available data.

We infer an occurrence rate of giant planets orbiting M dwarfs with periods up to two years in the range between 2\,\% to 6\,\%. The completion of the CARMENES survey, together with the release of astrometric orbit fits from $Gaia$, will place tighter bounds on this number. The combination of information from RV surveys and $Gaia$ will further improve our understanding of the architectures of M star systems, as the former are more sensitive to close-in planets and the latter to outer companions.

\begin{acknowledgements}

CARMENES is an instrument for the Centro Astron\'omico Hispano-Alem\'an (CAHA) at Calar Alto (Almer\'{\i}a, Spain), operated jointly by the Junta de Andaluc\'ia and the Instituto de Astrof\'isica de Andaluc\'ia (CSIC).

CARMENES was funded by the Max-Planck-Gesellschaft (MPG), the Consejo Superior de Investigaciones Cient\'{\i}ficas (CSIC), the Ministerio de Econom\'ia y Competitividad (MINECO) and the European Regional Development Fund (ERDF) through projects FICTS-2011-02, ICTS-2017-07-CAHA-4, and CAHA16-CE-3978, and the members of the CARMENES Consortium (Max-Planck-Institut f\"ur Astronomie, Instituto de Astrof\'{\i}sica de Andaluc\'{\i}a, Landessternwarte K\"onigstuhl, Institut de Ci\`encies de l'Espai, Institut f\"ur Astrophysik G\"ottingen, Universidad Complutense de Madrid, Th\"uringer Landessternwarte Tautenburg, Instituto de Astrof\'{\i}sica de Canarias, Hamburger Sternwarte, Centro de Astrobiolog\'{\i}a and Centro Astron\'omico Hispano-Alem\'an), with additional contributions by the MINECO, the Deutsche Forschungsgemeinschaft through the Major Research Instrumentation Program and Research Unit FOR2544 ``Blue Planets around Red Stars'', the Klaus Tschira Stiftung, the states of Baden-W\"urttemberg and Niedersachsen, and by the Junta de Andaluc\'{\i}a. We acknowledge financial support from NASA through grant NNX17AG24G. We acknowledge financial support from the Agencia Estatal de Investigaci\'on of the Ministerio de Ciencia, Innovaci\'on y Universidades through projects PID2019-110689RB-100, PID2019-109522GB-C5[1:4], PID2019-107061GB-C64 and the Centre of Excellence ``Severo Ochoa'', Instituto de Astrof\'{\i}sica de Andaluc\'{\i}a (SEV-2017-0709). We further acknowledge support by the BNSF program ``VIHREN-2021'' project No.\ K{\cyr P}-06-{\cyr D}B/5.

Based on data from the CARMENES data archive at CAB (CSIC-INTA).

Data were partly collected with the 150\,cm and 90\,cm telescopes at the Observatorio de Sierra Nevada (OSN) operated by the Instituto de Astrof\'{\i}sica de Andaluc\'{\i}a (IAA-CSIC).

This work makes use of observations from the Las Cumbres Observatory global telescope network.

This research has made use of the NASA Exoplanet Archive, which is operated by the California Institute of Technology, under contract with the National Aeronautics and Space Administration under the Exoplanet Exploration Program.

We thank the referee, Alexis Heitzmann, for his careful reading of the manuscript and his helpful comments.

\end{acknowledgements}

\bibliographystyle{aa}
{\tiny
\bibliography{Saturns}
}

\begin{appendix}

\section{Periodograms of spectroscopic indicators}
\label{Indicators}

\begin{figure}[h]
   \centering
   \includegraphics[width=\hsize]{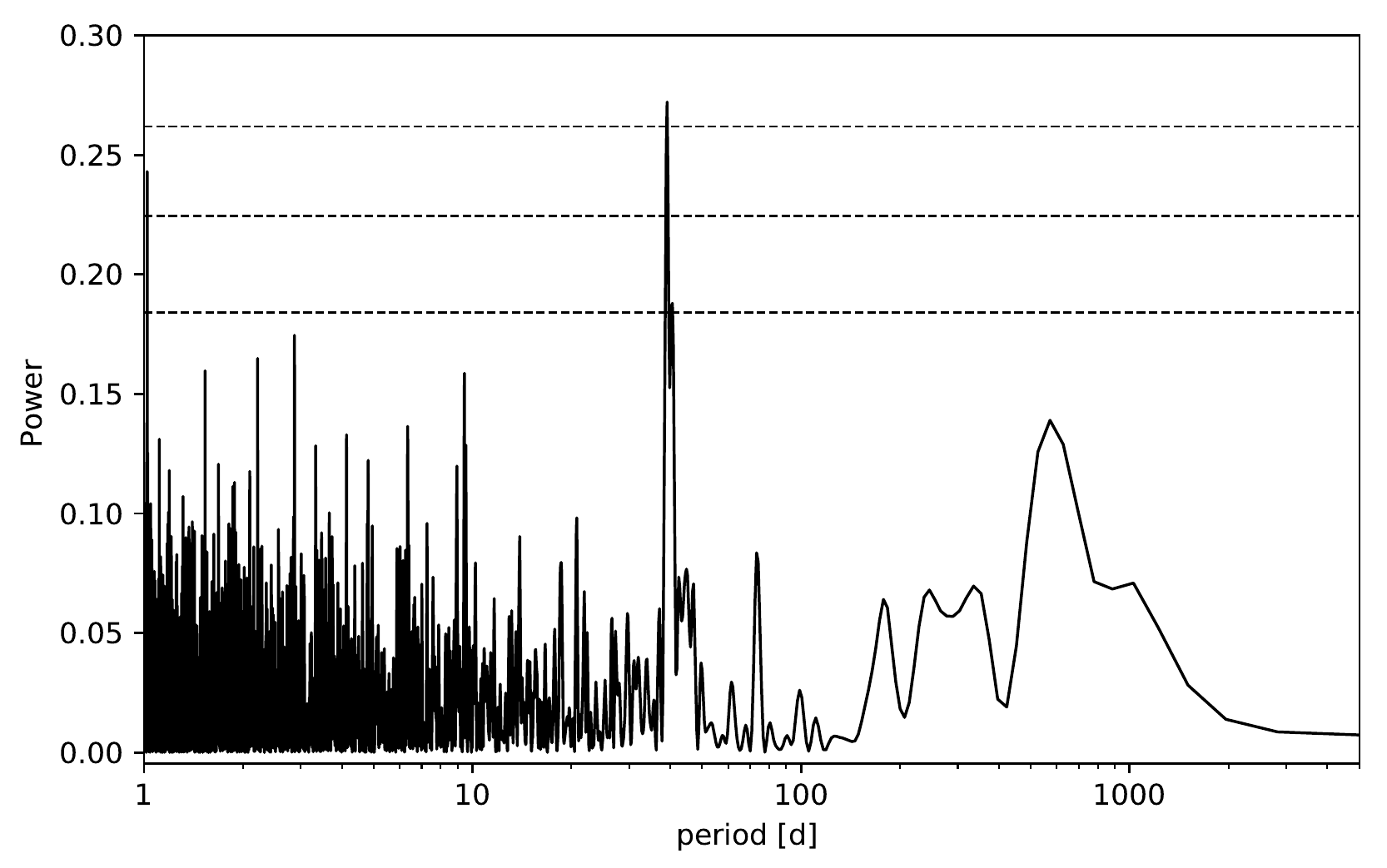}
      \caption{GLS periodogram of the dLW time series of \tyc{} from CARMENES. The prominent peak at 39.1\,d indicates the rotation period. False-alarm probabilities of 0.1, 0.01, and 0.001 are shown as dashed horizontal lines.}
         \label{Fig_TYC_dLW_GLS}
   \end{figure}

\begin{figure}[h]
   \centering
   \includegraphics[width=\hsize]{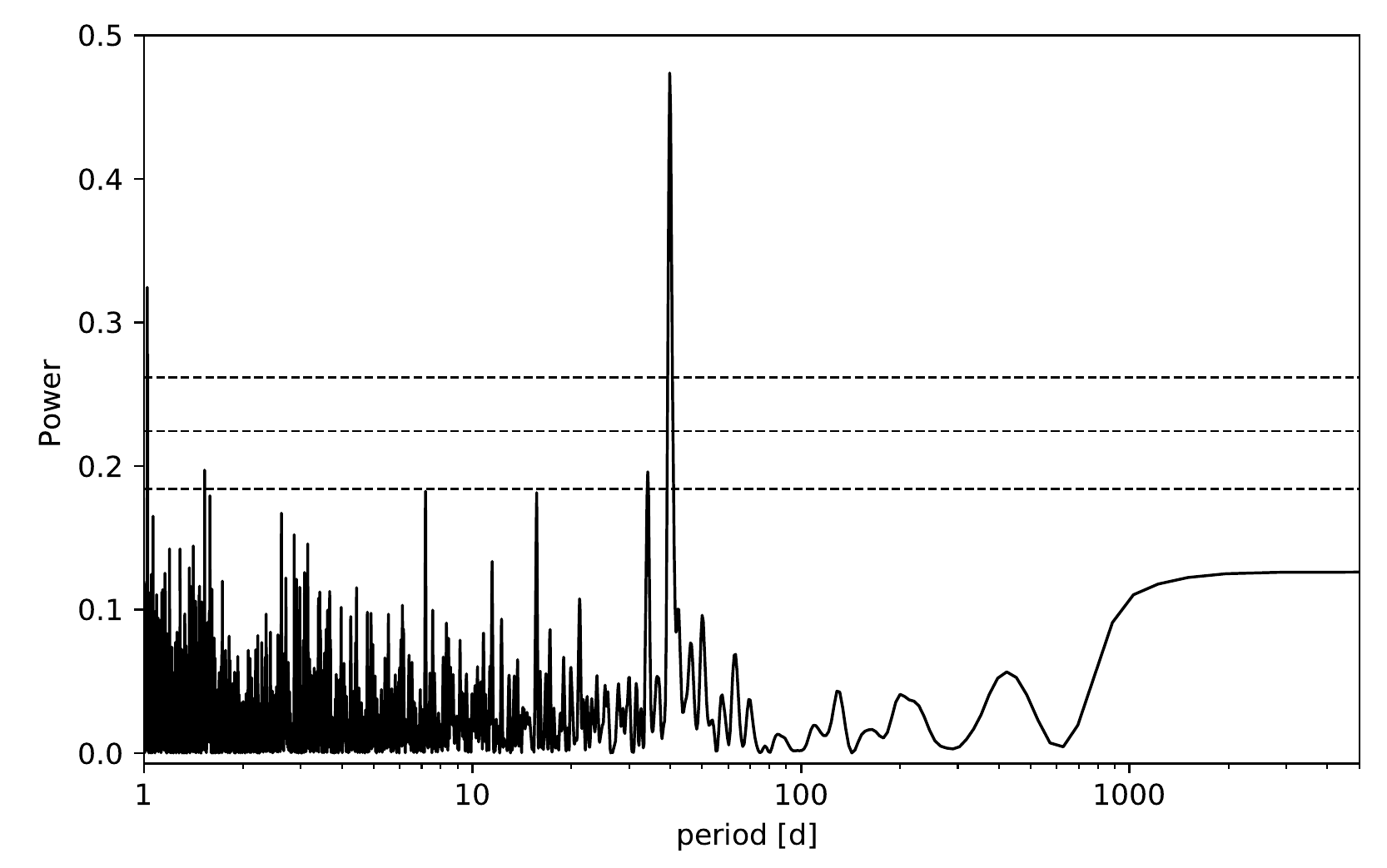}
      \caption{GLS periodogram of the H$\alpha$ line index time series of \tyc{} from CARMENES. The prominent peak at 39.9\,d indicates the rotation period. False-alarm probabilities of 0.1, 0.01, and 0.001 are shown as dashed horizontal lines.}
         \label{Fig_TYC_Halpha_GLS}
   \end{figure}

\begin{figure}
   \centering
   \includegraphics[width=\hsize]{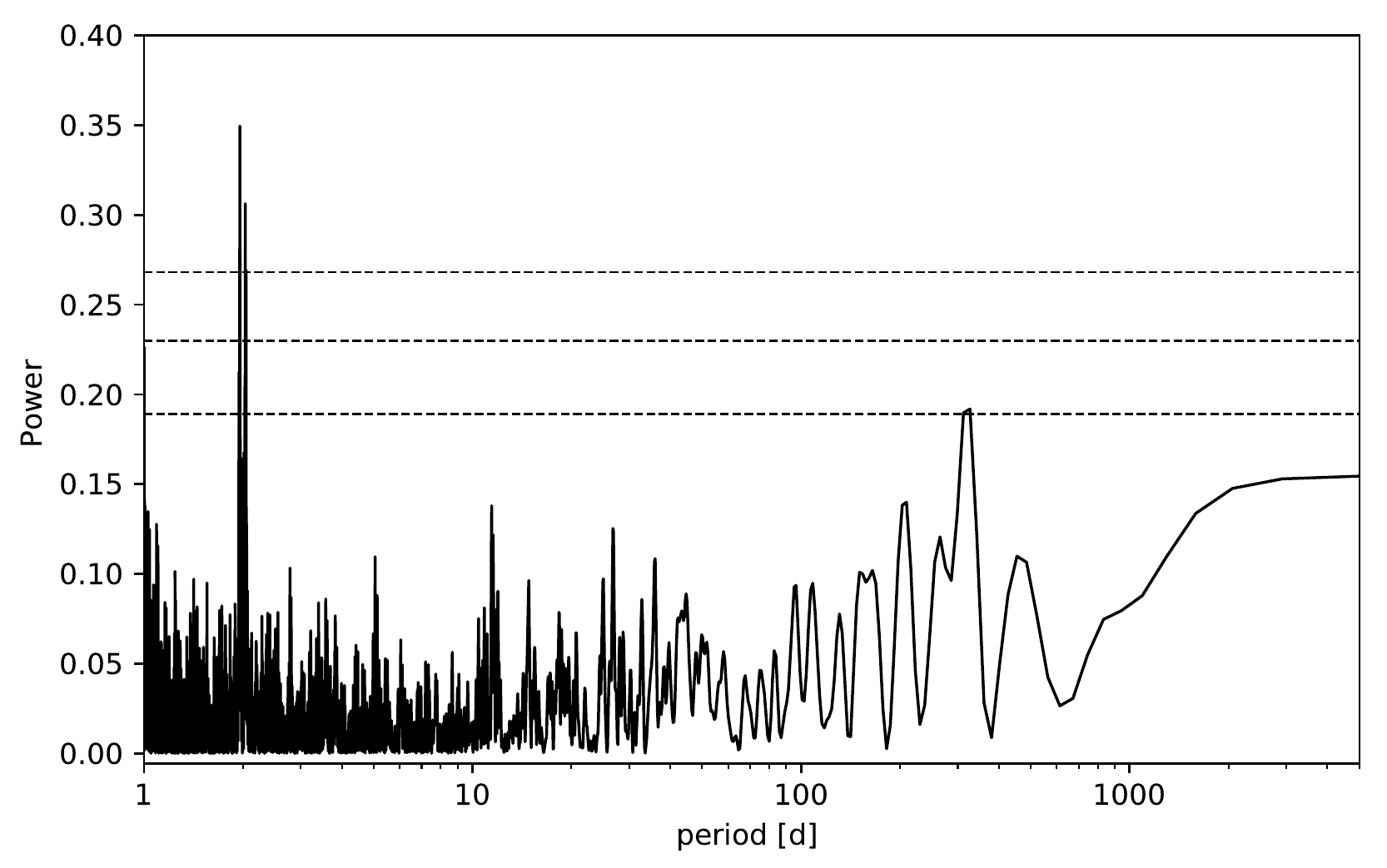}
      \caption{GLS periodogram of the CRX time series of TZ\,Ari from CARMENES. The prominent peak at 1.96\,d indicates the rotation period. False-alarm probabilities of 0.1, 0.01, and 0.001 are shown as dashed horizontal lines.}
         \label{Fig_TZ_CRX_GLS}
   \end{figure}

\begin{figure}
   \centering
   \includegraphics[width=\hsize]{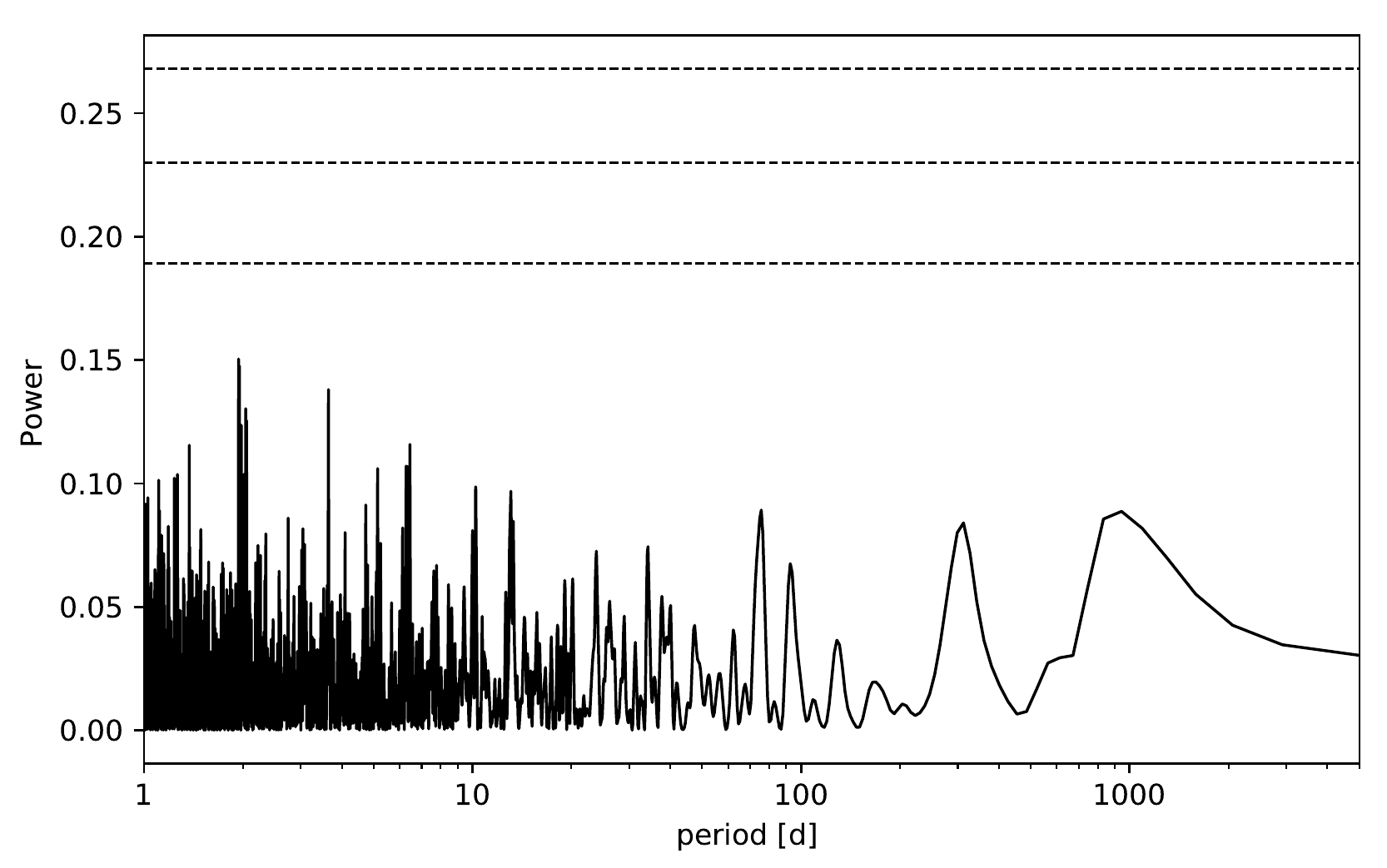}
      \caption{GLS periodogram of the H$\alpha$ line index time series of TZ\,Ari from CARMENES. The strongest peak at 1.94\,d is not significant, but it corresponds to the rotation period inferred from the other diagnostics. False-alarm probabilities of 0.1, 0.01, and 0.001 are shown as dashed horizontal lines.}
         \label{Fig_TZ_Halpha_GLS}
   \end{figure}

\begin{figure}
   \centering
   \includegraphics[width=\hsize]{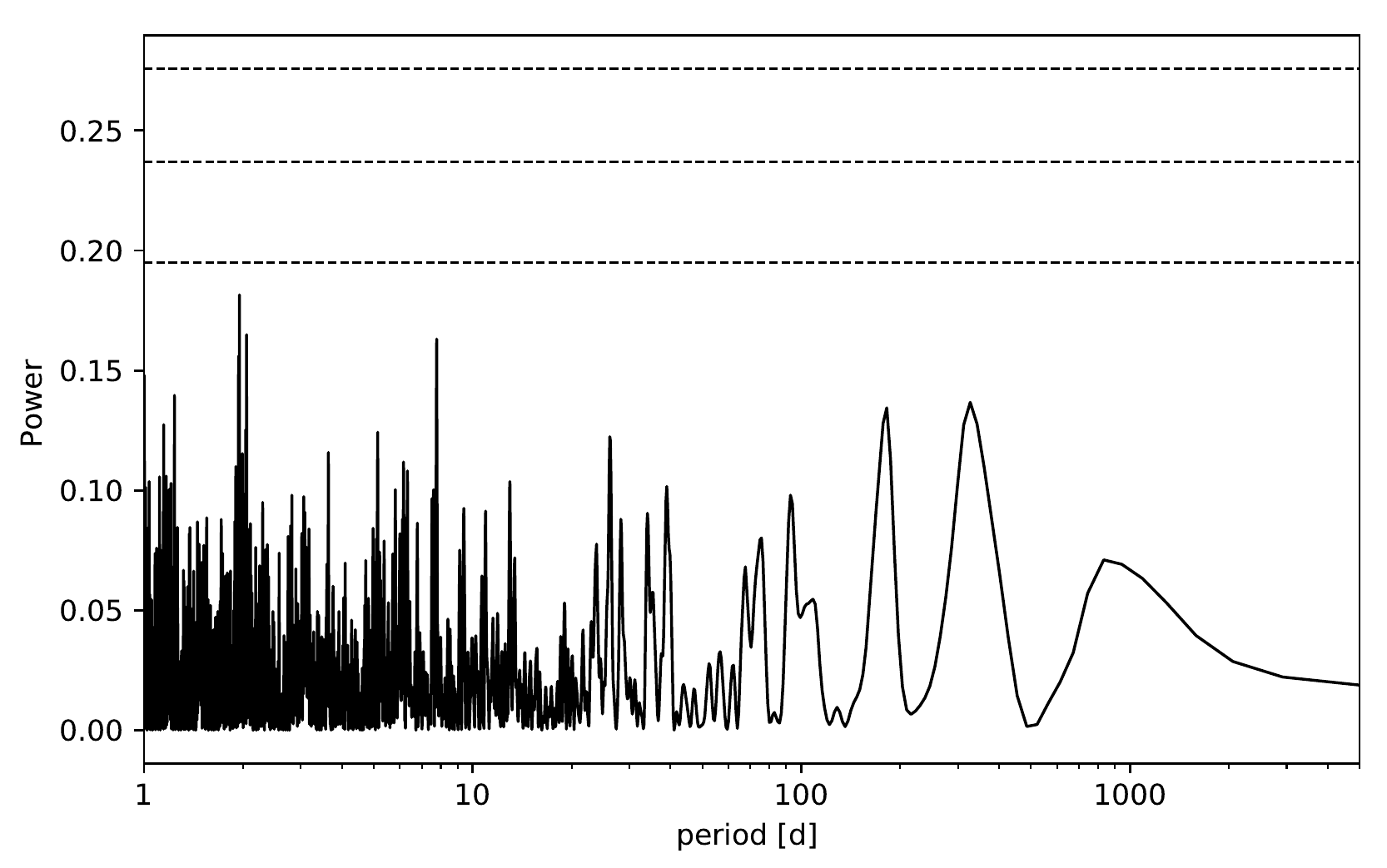}
      \caption{GLS periodogram of an average of the three Ca\,{\sc ii} IRT line index time series of TZ\,Ari from CARMENES. The strongest peak at 1.95\,d is not significant, but it corresponds to the rotation period inferred from the other diagnostics. False-alarm probabilities of 0.1, 0.01, and 0.001 are shown as dashed horizontal lines.}
         \label{Fig_TZ_IRT_GLS}
   \end{figure}

\clearpage

\section{Gaussian process priors}
\label{GP_priors}

\begin{table}[ht]
\centering
\caption{Priors adopted for the Gaussian process model fit to the RV data of \tyc{} from CARMENES VIS shown in Fig.~\ref{Fig_TYC_GP_RV} and in the right column of Table~\ref{tab:TYC_RVfits}.}
\label{tab:TYC_GP_Priors}

\begin{tabular}{lr}     

\hline\hline  \noalign{\vskip 0.7mm}
Parameter \hspace{0.0 mm}& GP prior \\
\hline \noalign{\vskip 0.7mm}

    $K$ [m\,s$^{-1}$]             & $\mathcal{U}(    10.000,    15.000)$\\
    $P$ [d]                       & $\mathcal{U}(   670.000,   730.000)$\\
    $e$                           & $\mathcal{U}(     0.000,     0.300)$\\
    $\omega$ [deg]                & $\mathcal{U}(  -720.000,   720.000)$\\
    $M_{\rm 0}$ [deg]             & $\mathcal{U}(  -720.000,   720.000)$\\
    RV$_{\rm off}$ [m\,s$^{-1}$]  & $\mathcal{U}(   -10.000,    10.000)$\\
    RV$_{\rm jit}$ [m\,s$^{-1}$]  & $\mathcal{U}(     0.000,    10.000)$\\
    GP$_{\rm SHO}$ $S$ [m$^{2}$\,s$^{-2}$\,d] & $\mathcal{N}(     2.000,     1.000)$\\
    GP$_{\rm SHO}$ $Q$            & $\mathcal{U}(     5.000,   100.000)$\\
    GP$_{\rm SHO}$ $\omega_0$ [d$^{-1}$] & $\mathcal{N}(     0.158,     0.005)$\\
\hline \noalign{\vskip 0.7mm}

\end{tabular}

\end{table}

\begin{table}[ht]
\centering
\caption{Priors adopted for the Gaussian process model fit to the RV data of TZ\,Ari from HIRES, CARMENES VIS, and HARPS (labeled as 1, 2, and 3, respectively) shown in Fig.~\ref{Fig_TZ_GP_RV} and in the right column of Table~\ref{tab:TZ_RVfits}.}
\label{tab:TZ_GP_Priors}

\begin{tabular}{lr}     

\hline\hline  \noalign{\vskip 0.7mm}
Parameter \hspace{0.0 mm}& GP prior \\
\hline \noalign{\vskip 0.7mm}

    $K$ [m\,s$^{-1}$]             & $\mathcal{U}(    11.000,    31.000)$\\
    $P$ [d]                       & $\mathcal{U}(   720.000,   820.000)$\\
    $e$                           & $\mathcal{U}(     0.000,     0.999)$\\
    $\omega$ [deg]                & $\mathcal{U}(  -720.000,   720.000)$\\
    $M_{\rm 0}$ [deg]             & $\mathcal{U}(  -720.000,   720.000)$\\
    RV$_{\rm off}$ 1 [m\,s$^{-1}$]& $\mathcal{U}(   -20.000,    20.000)$\\
    RV$_{\rm off}$ 2 [m\,s$^{-1}$]& $\mathcal{U}(   -20.000,    20.000)$\\
    RV$_{\rm off}$ 3 [m\,s$^{-1}$]& $\mathcal{U}(   -20.000,    20.000)$\\
    RV$_{\rm jit}$ 1 [m\,s$^{-1}$]& $\mathcal{U}(     0.000,    20.000)$\\
    RV$_{\rm jit}$ 2 [m\,s$^{-1}$]& $\mathcal{U}(     0.000,    20.000)$\\
    RV$_{\rm jit}$ 3 [m\,s$^{-1}$]& $\mathcal{U}(     0.000,    20.000)$\\
    GP$_{\rm SHO}$ $S$ [m$^{2}$\,s$^{-2}$\,d] & $\mathcal{N}(     0.010,     0.005)$\\
    GP$_{\rm SHO}$ $Q$         & $\mathcal{U}(     5.000,  1000.000)$\\
    GP$_{\rm SHO}$ $\omega_0$ [d$^{-1}$] & $\mathcal{N}(     3.200,     0.100)$\\
\hline \noalign{\vskip 0.7mm}

\end{tabular}

\end{table}

\onecolumn
\section{Corner plots}
\label{cornerplots}

\begin{figure*}[h]
   \centering
   \includegraphics[width=\hsize]{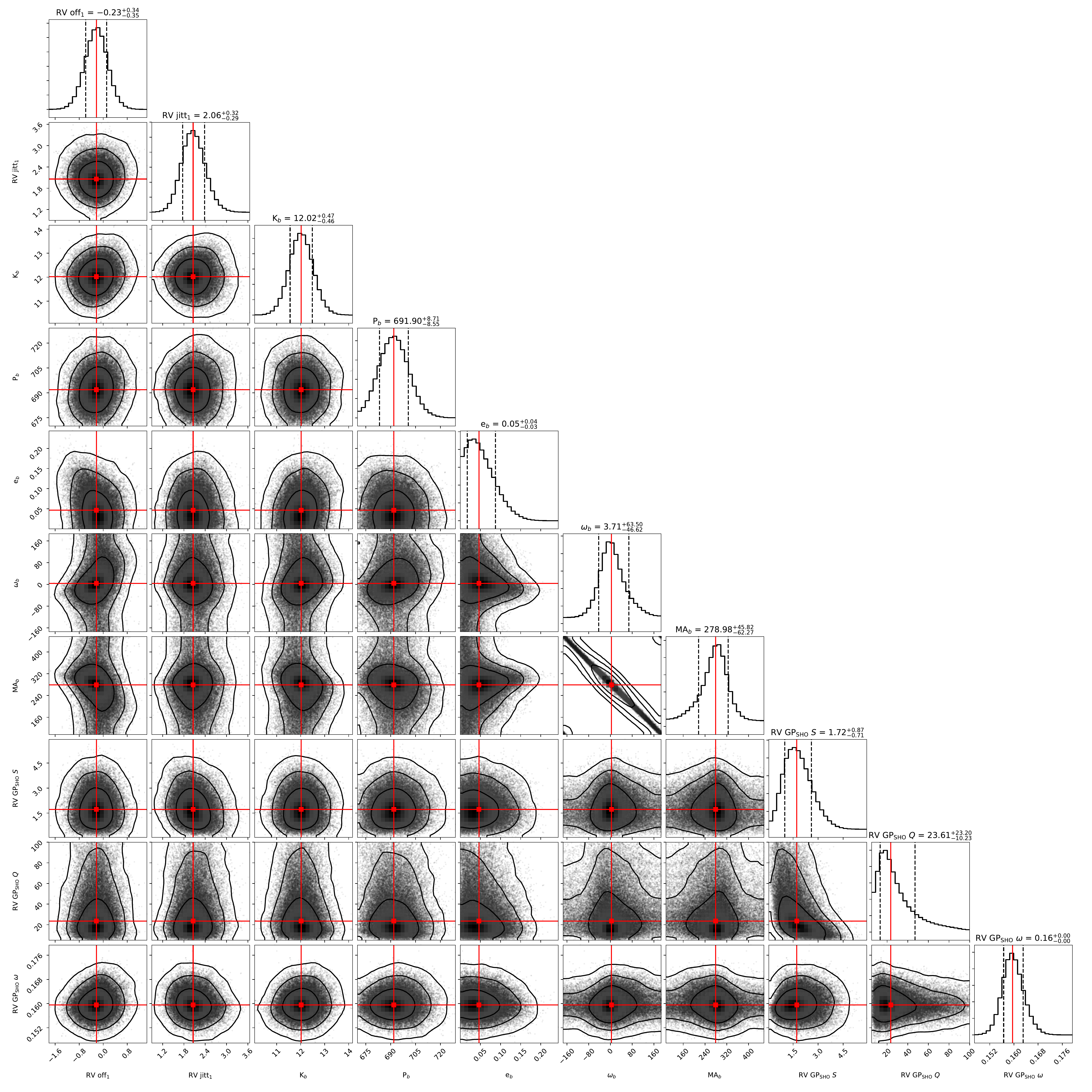}
      \caption{Corner plot for the Gaussian process model fit to the RV data of \tyc{} from CARMENES VIS shown in Fig.~\ref{Fig_TYC_GP_RV} and in the right column of Table~\ref{tab:TYC_RVfits}.}
         \label{Fig_TYC_GP_corner}
   \end{figure*}

\begin{figure*}[ht]
   \centering
   \includegraphics[width=\hsize]{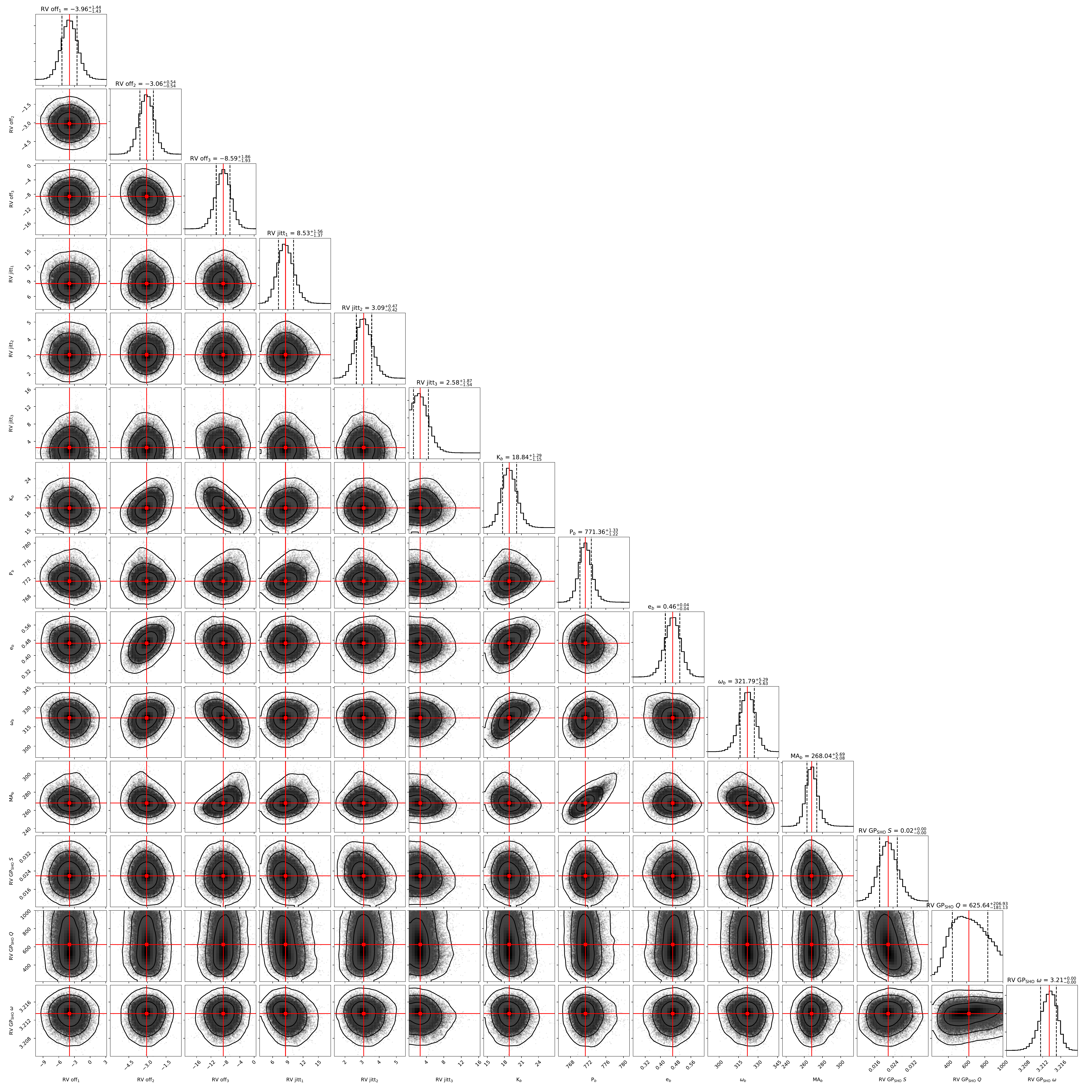}
      \caption{Corner plot for the Gaussian process model fit to the RV data of TZ\,Ari from HIRES, CARMENES VIS, and HARPS shown in Fig.~\ref{Fig_TZ_GP_RV} and in the right column of Table~\ref{tab:TZ_RVfits}.}
         \label{Fig_TZ_GP_corner}
   \end{figure*}

\end{appendix}

\end{document}